\documentclass[a4paper,11pt]{article}
\usepackage{jheppub}
\usepackage[T1]{fontenc}
\usepackage{stmaryrd}
\usepackage{amssymb,amsmath,amsfonts,nccmath,amsthm}
\usepackage{hyperref}
\usepackage{IEEEtrantools}
\usepackage{float}
\usepackage{physics}
\usepackage{braket}
\usepackage{tensor}
\usepackage{simplewick}
\usepackage{graphicx}
\usepackage{xcolor}
\usepackage{tensor}
\usepackage[utf8]{inputenc}
\usepackage{caption,subfig}
\usepackage{bm,bbm,bbold}	
\usepackage{tikz}	
\usepackage{scalerel}[2016/12/29]
\usepackage{enumitem}
\usepackage{multirow}
\usepackage{tabularx}
\usepackage{array, makecell}
\usepackage{mathtools}

\usetikzlibrary{arrows,positioning,decorations.markings,decorations.pathmorphing,calc}

\def\d{\mathrm{d}}

\def\E#1#{\tensor#1{\epsilon}}

\def\Q{\mathbb{Q}}

\newcommand{\red}[1]{{\color{red} #1}}

\def \be {\begin{equation}}
\def \ee {\end{equation}}

\def\D{\mathcal{D}}
\def\S{\mathcal{S}}
\def\T{\mathbb{T}}
\def\M{\mathcal{M}}

\def\C{\mathcal{C}}
\def\R{\mathcal{R}}

\newcommand{\G}[2]{\Gamma^{#1}{}_{#2}}
\newcommand{\ud}[2]{^{#1}{}_{#2}}
\newcommand{\du}[2]{_{#1}{}^{#2}}
\newcommand{\dud}[3]{_{#1}{}^{#2}{}_{#3}}

\newcommand{\transpose}{\intercal}
\newcommand{\ce}{\coloneqq}
\newcommand{\ec}{\eqqcolon}

\DeclareRobustCommand{\rchi}{{\mathpalette\irchi\relax}}
\newcommand{\irchi}[2]{\raisebox{\depth}{$#1\chi$}} 
\newcommand{\PD}[2]{\ensuremath{\frac{\partial #1}{\partial #2}}}

\newcommand{\varD}[2]{\ensuremath{\frac{\delta #1}{\delta #2}}}

\newcommand{\Hp}{H_\textsf{P}}

\title{Hamiltonian Analysis of $f(\Q)$ Gravity and the Failure of the Dirac--Bergmann Algorithm for Teleparallel Theories of Gravity}
\author[a,c]{Fabio D'Ambrosio,}
\author[a,b,c]{Lavinia Heisenberg,}
\author[a,c]{and Stefan Zentarra}
\affiliation[a]{Institute for Theoretical Physics,
ETH Z\"urich, Wolfgang-Pauli-Strasse 27, 8093, Z\"urich, Switzerland}
\affiliation[b]{Institut f\"{u}r Theoretische Physik, Philosophenweg 16, 69120 Heidelberg, Germany}
\affiliation[c]{Perimeter Institute, 31 Caroline Street N, Waterloo ON, Canada}
\emailAdd{fabioda@phys.ethz.ch}
\emailAdd{lavinia.heisenberg@phys.ethz.ch}
\emailAdd{szentarra@phys.ethz.ch}
\abstract{
 In recent years, $f(\Q)$ gravity has enjoyed considerable attention in the literature and important results have been obtained. However, the question of how many physical degrees of freedom the theory propagates---and how this number may depend on the form of the function $f$---has not been answered satisfactorily. In this article we show that a Hamiltonian analysis based on the Dirac-Bergmann algorithm---one of the standard methods to address this type of question---fails. We isolate the source of the failure, show that other commonly considered teleparallel theories of gravity are affected by the same problem, and we point out that the number of degrees of freedom obtained in~\href{https://journals.aps.org/prd/abstract/10.1103/PhysRevD.106.044025}{\textit{Phys. Rev.} \textbf{D} 106 no. 4, (2022)} by K.~Hu, T.~Katsuragawa, and T.~Qui (namely eight), based on the Dirac-Bergmann algorithm, is wrong. Using a different approach, we show that the upper bound on the degrees of freedom is seven. Finally, we propose a more promising strategy for settling this important question.
}


\begin{document}
	\allowdisplaybreaks[1]
	\maketitle
	\flushbottom
	

\section{Introduction}\label{sec:Introduction}
Gravity and geometry have been inseparable companions since the inception of General Relativity (GR), a brilliant formulation by Einstein that links gravity to the curvature of spacetime. This connection between gravity and curvature has proven to be remarkably efficient, such that gravitational phenomena are now commonly recognized as a manifestation of curved spacetime. This geometrical formulation of gravity is made possible by the equivalence principle, which renders gravitational interaction independent of the specific type of matter, and suggests an intriguing link between gravity and inertia. As a result, particle motion can naturally be associated with the geometrical properties of spacetime.

Embracing the geometrical nature of gravity, requires exploring different equivalent representations. At this point, it is necessary to recall that a manifold can be endowed with both a metric $g_{\mu\nu}$ and an affine connection $\Gamma\ud{\alpha}{\mu\nu}$. In a general metric-affine geometry, the metric and the connection are two entirely independent objects, which enable us to classify metric-affine geometries~\cite{Olmo:2011uz,Heisenberg:2018}. The failure of the connection to be metric-compatible is encoded in the non-metricity tensor, while its antisymmetric part defines the torsion. Within the set of all conceivable connections that may be introduced on a manifold, the Levi-Civita connection stands out as the unique connection that is both metric-compatible and symmetric. These two constraints unequivocally determine the Levi-Civita connection as being composed of the Christoffel symbols constructed from the metric tensor.

Einstein's original formulation of GR postulated a spacetime that was metric-compatible and devoid of torsion, and thus attributed gravity to the curvature of the said spacetime. Despite this, it was only natural for Einstein, as well as other researchers, to explore the possibility of assigning gravity to the remaining properties that spacetime may possess, namely torsion and non-metricity. It has been explicitly demonstrated that these three seemingly disparate elements, namely curvature, torsion, and non-metricity, can be used interchangeably to describe the same theory of gravity, namely GR. In doing so, a geometrical trinity of gravity, elucidating the unity between these three geometric elements of spacetime, was established~\cite{Heisenberg:2018,BeltranJimenez:2019}.

The groundwork laid by these different but equivalent geometric representations of GR provide a promising avenue for exploring modified theories of gravity. The equivalent depictions of GR that incorporate curvature, torsion, and non-metricity present distinct alternative foundations for theories of modified gravity. This is particularly true when one considers that the corresponding Lagrangians can be readily elevated to arbitrary non-linear functions. In recent literature, there has been considerable focus on theories of modified gravity that utilize scalar functions of the non-metricity tensor, known as $f(\Q)$ theories~\cite{BeltranJimenez:2017b}, with very interesting cosmological~\cite{BeltranJimenez:2019tme, BeltranJimenez:2019, DAmbrosio:2020c, Bajardi:2020, Ayuso:2020, Frusciante:2021, Anagnostopoulos:2021, Atayde:2021, DAmbrosio:2021b, Capozziello:2022, Dimakis:2022, Esposito:2022} as well as astrophysical~\cite{Zhao:2021, Lin:2021, DAmbrosio:2021, Banerjee:2021, Wang:2021, Parsaei:2022, Maurya:2022} implications.

Although $f(\Q)$ gravity has been subject of extensive research, certain aspects of this theory remain elusive. One such issue is the determination of its physical degrees of freedom. This is a matter of crucial importance as it directly affects the theory's cosmological and astrophysical perturbations and reveals any implicit symmetries that may have gone unnoticed. Determining these degrees of freedom was exactly the objective in~\cite{Hu:2022}, where  the Hamiltonian formulation and subsequently the Dirac-Bergmann procedure was used to analyze the constraints and obtain the number of propagating degrees of freedom. The conclusion was that $f(\Q)$ gravity exhibits precisely eight distinct physical degrees of freedom. 

In our investigation, it is revealed that a basic assumption of the Dirac-Bergmann algorithm is violated by $f(\Q)$ gravity, as well as other teleparallel gravity theories based on torsion or non-metricity, and that the analysis can therefore not be carried through. The deficiency is demonstrated explicitly in the case of $f(\Q)$ gravity, and the flaws in the analysis of~\cite{Hu:2022} are brought to attention. Moreover, we use a completely independent method to establish that $f(\Q)$ propagates at most seven physical degrees of freedom. 

The article is organized as follows: In section~\ref{sec:TeleparallelTheories} we review the basics of the Symmetric Teleparallel Equivalent of GR (STEGR). This serves the purpose of introducing basic concepts as well as fixing notations and conventions. Furthermore, we introduce two commonly studied extensions of STEGR: A five-parameter family of quadratic Lagrangians which define Symmetric Teleparallel Gravity (STG) (subsection~\ref{ssec:FiveParameterFamily}) and $f(\Q)$ gravity (subsection~\ref{ssec:f(Q)Gravity}). Subsection~\ref{ssec:TorsionTheories} briefly touches upon the main features of teleparallel theories of gravity based on torsion.  

Section~\ref{sec:DiracBergmannAlgorithm} is devoted to reviewing the basics of the Hamiltonian formalism and, in particular, the Dirac-Bergmann algorithm. This algorithm is summarized in subsection~\ref{ssec:SummaryAlgorithm}. In subsection~\ref{ssec:FailureAlgorithm} we show under which conditions the Dirac-Bergmann algorithm is \textit{not} applicable to field theories. The main part of this article are subsections~\ref{ssec:DiracBergmannAnalysis}, \ref{ssec:ObstaclesOtherTheories}, and section~\ref{sec:UpperBound}. Subsection~\ref{ssec:DiracBergmannAnalysis} is devoted to the Dirac-Bergmann analysis of $f(\Q)$ gravity. We explicitly show how and why the analysis fails and where the work of~\cite{Hu:2022} went wrong. Other teleparallel theories of gravity, based on either torsion or non-metricity, are scrutinized in subsection~\ref{ssec:ObstaclesOtherTheories}, where we argue that almost all teleparallel theories are affected by the same problem which leads to a failure of the Dirac-Bergmann approach. In section~\ref{sec:UpperBound} we employ an entirely different method to analyze $f(\Q)$ gravity and we establish that it propagates at least four and at most seven degrees of freedom. 

Finally, we draw our conclusions in section~\ref{sec:Conclusions} and we propose methods to circumvent the limitations of the Dirac-Bergmann algorithm. A complete analysis of $f(\Q)$ gravity and a determination of its physical degrees of freedom is left for future work.


\section{Teleparallel Theories of Gravity}\label{sec:TeleparallelTheories}
\subsection{The Symmetric Teleparallel Equivalent of GR}
The Symmetric Teleparallel Equivalent of General Relativity~\cite{Nester:1998, BeltranJimenez:2017b, BeltranJimenez:2018, Heisenberg:2018, BeltranJimenez:2019}, or STEGR for short, is a formulation of GR which is rooted in the mathematical framework of metric-affine geometry. That is, the effects of gravity are described by the triple $(\M, g, \Gamma)$, where~$\M$ is a real, connected, four-dimensional manifold, $g$ is a metric tensor of Lorentzian signature, and~$\Gamma$ is an affine connection. Standard GR can also be described within this mathematical framework. The only difference between GR and STEGR are a different set of geometric postulates, which are imposed to restrict the form of $\Gamma$, and a different form of the action.

To make this more precise, we first recall that the purpose of a connection is to define a notion of parallel transport or, equivalently, define a notion of covariant differentiation on~$\M$. The action of the covariant derivative $\nabla$ on vector fields $V$ and $1$-forms $\omega$ is defined as
\begin{align}
	\nabla_\mu V^\nu &\ce  \partial_\mu V^\nu + \G{\nu}{\mu\lambda}\, V^\lambda\notag\\
	\nabla_\mu \omega_\nu &\ce \partial_\mu \omega_\nu - \G{\lambda}{\mu\nu} \omega_\lambda\,.
\end{align}
Notice that it is always necessary to \textit{choose} an affine connection in order to properly define the action of the covariant derivative operator $\nabla$ on vectors, $1$-forms, and ultimately on any type of tensor (density). 

Given an affine connection $\Gamma$, one can define three types of tensors which characterize metric-affine geometries $(\M, g, \Gamma)$. These tensors are
\begin{align}
	&\text{Curvature tensor:} & R\ud{\alpha}{\mu\nu\rho} &\ce 2\partial_{[\nu}\G{\alpha}{\rho]\mu} + 2 \G{\alpha}{[\nu|\lambda}\G{\lambda}{\rho]\mu}\notag\\
	&\text{Torsion tensor:} & T\ud{\alpha}{\mu\nu} &\ce 2\G{\alpha}{[\mu\nu]}\notag\\
	&\text{Non-metricity tensor:} & Q_{\alpha\mu\nu} &\ce \nabla_\alpha g_{\mu\nu} = \partial_\alpha g_{\mu\nu} - 2\G{\lambda}{\alpha(\mu}g_{\nu)\lambda}\, ,
\end{align}
where the round and square brackets symbolize symmetrization and anti-symmetrization of indices, respectively.   Metric-affine geometries can now be classified according to which of these tensors (or which combination of these tensors) is not zero. Riemannian geometry, which constitutes the mathematical framework of GR, emerges as special case of a metric-affine geometry, where the torsion and non-metricity tensors vanish. Observe that this amounts to restrictions on the choice of $\Gamma$, since the connection has to satisfy the equations $\Gamma_{[\mu\nu]} = 0$ and $\partial_\alpha g_{\mu\nu} = 2\G{\lambda}{\alpha(\mu}g_{\nu)\lambda}$. As is well-known, these conditions completely determine the connection, leaving us no free choice. In fact, the connection then has to be the Levi-Civita connection, which we denote by $\left\{\alpha\atop\mu\nu\right\}$ and which is explicitly given by
\begin{align}\label{eq:LeviCivitaConnection}
	\left\{\alpha\atop\mu\nu\right\} = \frac12 g^{\alpha\lambda}\left(\partial_{\mu} g_{\nu \lambda} + \partial_{\nu} g_{\mu\lambda}-\partial_\lambda g_{\mu\nu}\right)\,.
\end{align}
STEGR, however, is based on a different set of geometric postulates. Concretely, the connection is postulated to have vanishing curvature and vanishing torsion:
\begin{align}\label{eq:GeometricPostulates}
	R\ud{\alpha}{\mu\nu\rho} &\overset{!}{=} 0 & \text{and} && T\ud{\alpha}{\mu\nu} &\overset{!}{=} 0\, .
\end{align}
These postulates leave the non-metricity tensor $Q_{\alpha\mu\nu}$ as the only non-vanishing\footnote{If in addition to curvature and torsion also the non-metricity tensor vanished, one would be left with Minkowski spacetime. In order to consider non-trivial spacetimes, one needs $Q_{\alpha\mu\nu}\neq 0$ when curvature and torsion vanish.} tensor and the effects of gravity can be attributed to this tensor. Moreover, these postulates do not completely fix the connection. In fact, one can show~\cite{BeltranJimenez:2017b, BeltranJimenez:2018, Heisenberg:2018, BeltranJimenez:2019} that a connection which gives rise to a vanishing curvature tensor necessarily has the form
\begin{align}\label{eq:FlatConnection}
	\G{\alpha}{\mu\nu} = (\Lambda^{-1})\ud{\alpha}{\lambda}\partial_\mu\Lambda\ud{\lambda}{\nu}\, ,
\end{align}
where $\Lambda\ud{\mu}{\nu}$ are the components of a matrix which belongs to the general linear group $GL(4, \mathbb R)$, i.e., the set of invertible $4\times 4$ matrices. Using~\eqref{eq:FlatConnection}, the postulate of vanishing torsion takes the form
\begin{align}
	\partial_{[\mu}\Lambda\ud{\alpha}{\nu]} = 0\,.
\end{align}
This condition is solved by matrices of the form
\begin{equation}\label{eq:MatrixForm}
	\Lambda\ud{\alpha}{\nu} = \partial_\nu \xi^\alpha\,,
\end{equation}
where $\xi^\alpha$ denotes four arbitrary\footnote{The only condition one has to impose on these functions is that the determinant of the matrix $\partial_\nu \xi^\mu$ is not zero. That is because matrices belonging to $GL(4, \mathbb R)$ are invertible and thus have a non-zero determinant.} functions of the spacetime coordinates, \textit{not} a vector field! By combining~\eqref{eq:FlatConnection} and~\eqref{eq:MatrixForm}, one finds that a flat and torsionless connection has the form
\begin{align}\label{eq:FlatTorsionless}
	\G{\alpha}{\mu\nu} = \PD{x^\alpha}{\xi^\lambda} \partial_\mu \partial_\nu \xi^\lambda\,,
\end{align}
where $\partial x^\alpha/\partial\xi^\lambda$ should be read as the inverse of the Jacobian matrix $\partial_\alpha \xi^\lambda \ce \partial\xi^\lambda/\partial x^\alpha$. We emphasize that any affine connection of the form~\eqref{eq:FlatTorsionless} is guaranteed to be flat and torsionless. Moreover, this form of the connection uncovers a particular feature of gravity theories based on flat and torsionless metric-affine geometries: The connection can be trivialized globally by a particular gauge choice. In fact, by choosing $\xi^\alpha = M\ud{\alpha}{\lambda}\,x^\lambda + \xi^\alpha_0$, where $M\ud{\alpha}{\lambda}$ is an invertible matrix with constant components and $\xi^\alpha_0$ denotes four arbitrary constants, one finds
\begin{equation}
	\partial_\mu\partial_\nu \xi^\alpha = M\ud{\alpha}{\lambda}\partial_\mu\partial_\nu x^\lambda = 0\,.
\end{equation}
Thus, if the four functions $\xi^\alpha$ are chosen to have the form $\xi^\alpha = M\ud{\alpha}{\lambda}\,x^\lambda + \xi^\alpha_0$, the connection vanishes globally, $\G{\alpha}{\mu\nu} = 0$. This is known as the \textit{coincident gauge}~\cite{BeltranJimenez:2017b, BeltranJimenez:2018, Heisenberg:2018, BeltranJimenez:2019}. We also emphasize that the coincident gauge is a feature common to every theory based on a metric-affine geometry $(\M, g, \Gamma)$ with a flat and torsionless connection, independent of what the Lagrangian of the theory might be.

Having established the geometric foundations of symmetric teleparallel theories of gravity, we now turn to the action which defines STEGR and its dynamics. There are actually two action functionals one can consider. The first, which is based on a flat and torsionless connection of the form~\eqref{eq:FlatTorsionless}, is given by~\cite{BeltranJimenez:2017b, BeltranJimenez:2019}
\begin{align}\label{eq:ActionSTEGR}
	\S_\text{STEGR}[g, \xi] \ce -\frac{1}{16\pi G} \int_{\M}\d^4 x\, \sqrt{|g|}\, \Q(g, \xi) + \S_\text{matter}\,,
\end{align}
where $g$ denotes the determinant of the metric, $\S_\text{matter}$ stands for the action of minimally coupled matter fields, and $\Q$ is the so-called non-metricity scalar,
\begin{equation}\label{eq:NonMetricityScalar}
	\Q \ce \frac14 Q_{\alpha\mu\nu}Q^{\alpha\mu\nu} - \frac12 Q_{\alpha\mu\nu}Q^{\mu\alpha\nu} - \frac14 Q_\alpha Q^\alpha + \frac12 Q_\alpha \bar{Q}^\alpha\, ,
\end{equation}
and where
\begin{align}
	Q_\alpha &\ce Q\du{\alpha\lambda}{\lambda} &&\text{and}& \bar{Q}_\alpha &\ce Q\ud{\lambda}{\lambda\alpha}
\end{align}
are the two independent traces of the non-metricity tensor. 
Observe that the action~\eqref{eq:ActionSTEGR} is a functional of the metric $g_{\mu\nu}$ and the four functions $\xi^\alpha$. A different action functional, based on the Palatini principle, which regards the metric and the connection as two independent fields, is given by~\cite{BeltranJimenez:2017b, BeltranJimenez:2018}
\begin{equation}\label{eq:FullyCovariantAction}
	\S_\text{STEGR}[g, \Gamma; \tilde{\Pi}, \tilde{\rchi}] \ce -\frac{1}{16\pi G}\int_{\M}\d^4x\,  \left(\sqrt{|g|}\,\Q(g, \Gamma) + \tilde{\Pi}\du{\alpha}{\mu\nu\rho} R\ud{\alpha}{\mu\nu\rho} + \tilde{\rchi}\du{\alpha}{\mu\nu} T\ud{\alpha}{\mu\nu}\right)\,,
\end{equation}
where $\tilde{\Pi}\du{\alpha}{\mu\nu\rho}$ and $\tilde{\rchi}\du{\alpha}{\mu\nu}$ are tensor densities\footnote{Our convention is to put a tilde $\tilde{\phantom{.}}$ on all tensor densities of weight $w=+1$, except the Lagrangian and Hamiltonian densities, which we denote by $\mathcal L$ and $\mathcal H$, respectively.} of weight $w=+1$. These tensor densities act as Lagrange multipliers and by varying the action~\eqref{eq:FullyCovariantAction} with respect to these multipliers, one obtains the equations
\begin{align}\label{eq:ConstraintsGamma}
	\frac{\delta\S_\text{STEGR}}{\delta \tilde{\Pi}\du{\alpha}{\mu\nu\rho}} &\overset{!}{=} 0 && \Longrightarrow &R\ud{\alpha}{\mu\nu\rho} &\overset{!}{=} 0\notag\\
	\frac{\delta\S_\text{STEGR}}{\delta \tilde{\rchi}\du{\alpha}{\mu\nu}} &\overset{!}{=} 0 && \Longrightarrow &T\ud{\alpha}{\mu\nu} &\overset{!}{=} 0\,.
\end{align}
Thus, the difference between the actions~\eqref{eq:ActionSTEGR} and~\eqref{eq:FullyCovariantAction} is that in~\eqref{eq:ActionSTEGR} one uses the parametrization~\eqref{eq:FlatTorsionless}, which directly implements the postulates of vanishing curvature and vanishing torsion, whereas in~\eqref{eq:FullyCovariantAction} the connection $\G{\alpha}{\mu\nu}$ is a general affine-connection. Only after varying with respect to the Lagrange multipliers and solving the equations~\eqref{eq:ConstraintsGamma} does one obtain a flat and torsionless connection. Ultimately, the actions~\eqref{eq:ActionSTEGR} and~\eqref{eq:FullyCovariantAction} define the same theory. Namely, General Relativity. 

One can show~\cite{BeltranJimenez:2017b, Heisenberg:2018, BeltranJimenez:2019} that the action~\eqref{eq:ActionSTEGR} is equal to the Einstein-Hilbert action of GR, up to a boundary term. This result follows from the geometric identity
\begin{equation}\label{eq:RicciQId}
	\R = -\Q -\D_\alpha\left(Q^\alpha - \bar{Q}^\alpha\right)\, ,
\end{equation}
where $\R$ and $\D_\alpha$ are the Ricci scalar and covariant derivative operator with respect to the Levi-Civita connection~\eqref{eq:LeviCivitaConnection} (not with respect to the affine connection $\Gamma$). From this identity it follows at once that STEGR and GR give rise to the same field equations for the metric. It is therefore also no surprise that GR and STEGR are equivalent and propagate the same number of physical degrees of freedom: two. This fact has also been checked by an explicit Hamiltonian analysis of the STEGR action~\eqref{eq:ActionSTEGR} in the coincident gauge~\cite{DAmbrosio:2020}.

 However, flat and torsionless metric-affine geometries can also be taken as starting point for constructing modified theories of gravity. There are two commonly considered modifications in the literature and for these theories it is in general no longer true that there are two physical degrees of freedom. In the following two subsections we define these two theories and discuss some of their properties. 

\subsection{Five-Parameter Family of Quadratic Non-Metricity Lagrangians}\label{ssec:FiveParameterFamily}
In the previous subsection, we saw that the action of STEGR is based on the non-metricity scalar. This scalar is quadratic in the non-metricity tensor and one can naturally ask, what is the most general scalar quantity one can construct which is quadratic in the non-metricity tensor? Given that the non-metricity tensor is symmetric in its last two indices, one finds that there are only five terms which can be constructed from quadratic contractions of the non-metricity tensor~\cite{BeltranJimenez:2017b, Heisenberg:2018, BeltranJimenez:2018, BeltranJimenez:2019, DAmbrosio:2020b}. Thus, the most general non-metricity tensor one can consider reads
\begin{align}\label{eq:GeneralNonMetricityScalar}
	\Q = c_1\, Q_{\alpha\mu\nu}Q^{\alpha\mu\nu} + c_2\, Q_{\alpha\mu\nu} Q^{\mu\alpha\nu} + c_3\, Q_\alpha Q^\alpha + c_4\, \bar{Q}_\alpha \bar{Q}^\alpha + c_5\, Q_\alpha\bar{Q}^\alpha\,,
\end{align} 
where $c_1$, $c_2$, $c_3$, $c_4$, and $c_5$ are real but otherwise arbitrary constants. The non-metricity scalar~\eqref{eq:GeneralNonMetricityScalar} gives rise to a five-parameter family of gravity theories, which we dub Symmetric Teleparallel Gravity (STG) theories, described by the action functional
\begin{equation}\label{eq:QuadraticAction}
	\S_\text{STG}[g, \Gamma; \tilde{\Pi}, \tilde{\rchi}] \ce -\frac{1}{16\pi G}\int_{\M}\d^4x\,  \left(\sqrt{|g|}\,\Q(g, \Gamma; c_i) + \tilde{\Pi}\du{\alpha}{\mu\nu\rho} R\ud{\alpha}{\mu\nu\rho} + \tilde{\rchi}\du{\alpha}{\mu\nu} T\ud{\alpha}{\mu\nu}\right)\,.
\end{equation}
One can check that the choice 
\begin{align}\label{eq:STEGRParameters}
	c_1 &= -\frac14\, , & c_2 &= \frac12\, , & c_3 &= \frac14\,, & c_4 &= 0\,,&\text{and} & & c_5 &= -\frac12
\end{align}
reproduces the non-metricity scalar~\eqref{eq:NonMetricityScalar} and thus the STG action~\eqref{eq:QuadraticAction} reduces to the STEGR action~\eqref{eq:FullyCovariantAction}. Furthermore, it is important to realize that the geometric identity~\eqref{eq:RicciQId} holds \textit{only} for this choice of parameters, i.e., only for the choice which corresponds to STEGR. For any other choice of parameters, the most general non-metricity scalar~\eqref{eq:GeneralNonMetricityScalar} differs from the Ricci scalar $\R$ of the Levi-Civita connection by \textit{more} than just a boundary term. Thus, any STG theory based on a choice of parameters \textit{different} from the one of STEGR leads to a modification of GR. 

However, not every choice of parameters might lead to a viable theory. The field equations derived from~\eqref{eq:QuadraticAction} are clearly at most second order field equations for the metric, since the non-metricity tensor only contains first order derivatives of the same. This excludes instabilities due to higher derivatives, but it does not preclude the presence of ghosts. It is also possible to choose the parameters such that one obtains a non-trivial action but either one or zero propagating degrees of freedom. A detailed analysis of the five-parameter action~\eqref{eq:QuadraticAction} based on the coincident gauge (recall that the coincident gauge is common to all flat and torsionless metric-affine geometries, irrespective of any Lagrangian) was performed in~\cite{DAmbrosio:2020b}. 

\begin{figure}[hbt!]
	\centering
	\includegraphics[width=1\columnwidth]{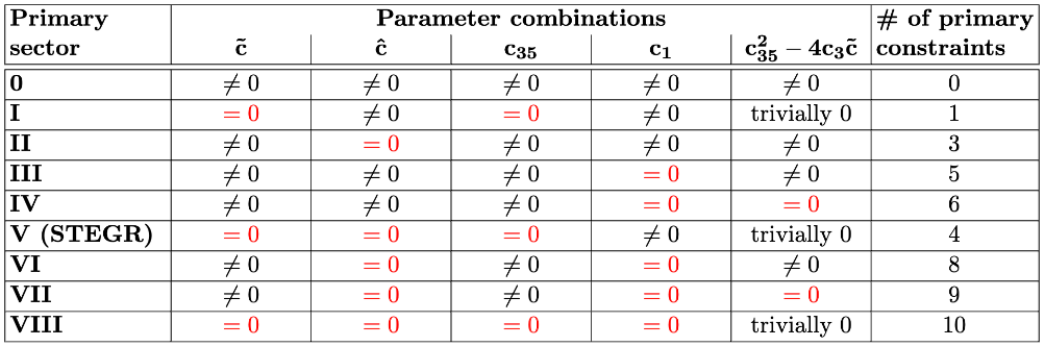}
	\caption{\protect \textit{Each primary sector (first column) is defined by the vanishing of certain combinations of the parameters $c_1$, $c_2$, $c_3$, $c_4$, and $c_5$ (columns two through six). For brevity, and following~\cite{DAmbrosio:2020b}, we have defined $\tilde{c}\ce c_1+c_2+c_3+c_4+c_5$, $\hat{c} \ce 2c_1 + c_2 + c_4$, and $c_{35} \ce 2c_3 + c_5$. The vanishing of these parameters (or the vanishing of combinations thereof) corresponds to the appearance of one or more primary constraints (the number of independent primary constraints is shown in the last column). STEGR is contained in sector V.}\hspace*{\fill}}
	\label{fig:PrimarySectors}
\end{figure}

The analysis was based on the first step of the Dirac-Bergmann algorithm, which consists in finding all so-called primary constraints (see subsection~\ref{ssec:SummaryAlgorithm} for a definition of this terminology). Primary constraints give us an upper bound on the number of propagating degrees of freedom and in~\cite{DAmbrosio:2020b} it was found that the number of primary constraints depends on how one chooses the parameters $c_1$, $c_2$, $c_3$, $c_4$, and $c_5$. It was found that the five-parameter space of theories compartmentalizes into nine different sectors, as summarized in Figure~\ref{fig:PrimarySectors}. Each sector is characterized by a different number of primary constraints, or, put differently, is characterized by a different upper bound on the number of propagating degrees of freedom. These insights can already be used to conclude that certain sectors in the five-parameter space can only contain unphysical theories. We refer the reader to~\cite{DAmbrosio:2020b} for more details on this.
For the remaining sectors it is not possible to draw any conclusions from the primary constraints alone. One therefore has to proceed to the second step of the Dirac-Bergmann algorithm, which is to determine the so-called secondary constraints, as we will review in subsection~\ref{ssec:SummaryAlgorithm}. This step has not yet been carried out and, as we will show in subsection~\ref{ssec:FailureAlgorithm}, it is unlikely that one will succeed. The problem is that the Dirac-Bergmann algorithm suffers from a little known shortcoming and unless one is lucky, it is not possible to carry out the second step of the algorithm for field theories (there is no obstruction for constrained Lagrangians describing point-particles).

\subsection{Non-Linear Extension: $f(\Q)$ Gravity}\label{ssec:f(Q)Gravity}
Another possible modification of gravity based on flat and torsionless metric-affine geometries is $f(\Q)$ gravity. This theory represents a non-linear extension of STEGR, where the non-metricity scalar~\eqref{eq:NonMetricityScalar} (not the scalar~\eqref{eq:GeneralNonMetricityScalar}) in the STEGR action~\eqref{eq:FullyCovariantAction} is replaced by $f(\Q)$, where $f$ is an arbitrary function solely subjected to the condition $\frac{\d f}{\d \Q} \neq 0$. This condition ensures that the resulting theory has a non-trivial dynamics, as one can determine from looking at the field equations for the metric derived from the action functional~\cite{BeltranJimenez:2017b, Heisenberg:2018, BeltranJimenez:2018}
\begin{align}\label{eq:f(Q)Action}
	\S_{f(\Q)}[g, \Gamma; \tilde{\Pi}, \tilde{\rchi}] \ce -\frac{1}{16\pi G}\int_{\M}\d^4x\,  \left(\sqrt{|g|}\,f(\Q) + \tilde{\Pi}\du{\alpha}{\mu\nu\rho} R\ud{\alpha}{\mu\nu\rho} + \tilde{\rchi}\du{\alpha}{\mu\nu} T\ud{\alpha}{\mu\nu}\right)\,.
\end{align}
This non-linear extension of STEGR has received considerable attention in the literature. In particular, the existence and properties of stationary and spherically symmetric solutions which can describe stars, black holes, or wormholes haven been studied~\cite{Zhao:2021, Lin:2021, DAmbrosio:2021, Banerjee:2021, Wang:2021, Parsaei:2022, Maurya:2022}. An even more active field of research is $f(\Q)$ cosmology~\cite{BeltranJimenez:2019tme, BeltranJimenez:2019, DAmbrosio:2020c, Bajardi:2020, Ayuso:2020, Frusciante:2021, Anagnostopoulos:2021, Atayde:2021, DAmbrosio:2021b, Capozziello:2022, Dimakis:2022, Esposito:2022}.

However, the question of how many degrees of freedom propagate in $f(\Q)$ gravity has received relatively little attention. In~\cite{DAmbrosio:2020}, the Hamiltonian analysis of STEGR in the coincident gauge was carried out and an educated guess about the number of degrees of freedom in $f(\Q)$ was attempted. The argument given in that reference is based on
\begin{itemize}
	\item[(a)] findings from cosmological perturbation theory obtained in~\cite{BeltranJimenez:2019tme}, where it was found that $f(\Q)$ propagates \textit{at least} two additional degrees of freedom compared to GR;
	\item[(b)] the expectation that the primary constraints of $f(\Q)$ are all second class, since the theory is generally covariant.
\end{itemize}
Putting (a) and (b) together leads to the conclusion that $f(\Q)$ propagates six degrees of freedom. Of course, this argument is purely heuristic and needs to be followed up by a rigorous analysis of the theory.

Such an analysis has been carried out in~\cite{Hu:2022}, where the Dirac-Bergmann algorithm was employed~\cite{Dirac:1950, Bergmann:1951, DiracBook}. The algorithm allegedly determined the number of physical degrees of freedom to be eight, thereby challenging the guess ventured in~\cite{DAmbrosio:2020}. Moreover, the authors of~\cite{Hu:2022} argued that this large number has to be explained with the breaking of diffeomorphism symmetry of the theory. The reason for breaking this symmetry is the use of the coincident gauge during the analysis.

It seems that the Dirac-Bergmann analysis of~\cite{Hu:2022} supersedes the heuristic argument of~\cite{DAmbrosio:2020} and therefore settles the question. However, as we will show in the next section, the situation is not as clear-cut and simple as it might seem. In fact, we will point out an error in the analysis of~\cite{Hu:2022}, which will reopen the question. Unfortunately, even after identifying the error, we are not able to give a definitive answer regarding the number of physical degrees of freedom. Rather, we uncover a shortcoming of the Dirac-Bergmann algorithm which afflicts field theories. We explicitly show how this shortcoming prevents one from completing the Dirac-Bergmann analysis of $f(\Q)$ gravity and we show that many other teleparallel theories of gravity are potentially affected by this problem---this includes theories based on non-metricity as well as theories based on torsion. Therefore, before reviewing the Dirac-Bergmann algorithm and isolating the problem, we will briefly introduce torsion theories of gravity.

\subsection{Teleparallel Theories of Gravity based on Torsion}\label{ssec:TorsionTheories}
We have seen that metric-affine geometries provide the mathematical framework for two distinct, but ultimately equivalent formulations of gravity: GR and STEGR. These two formulations are distilled from the most general metric-affine geometry $(\M, g, \Gamma)$, where $\Gamma$ is a connection which gives rise to non-trivial curvature, torsion, and non-metricity tensors, by imposing certain postulates. In GR one demands that $\Gamma$ is chosen such that torsion and non-metricity vanish, which unambiguously fixes the connection to be the Levi-Civita connection. STEGR, in turn, is based on a flat and torsionless connection. 

By demanding a flat and metric-compatible connection with non-trivial torsion, one establishes the basis for a third formulation of gravity: The so-called Teleparallel Equivalent of GR (TEGR). This approach to gravity complements the other two and together GR, STEGR, and TEGR form what is known as the geometric trinity of GR~\cite{Heisenberg:2018,BeltranJimenez:2019}. 

Of course, the postulates of vanishing curvature and non-metricity do not yet fully define a theory. One has to stipulate an action principle, which in the  case of TEGR reads
\begin{align}\label{eq:TEGRAction}
	\S_\text{TEGR}[g, \Gamma; \tilde{\Pi}, \tilde{\Xi}] \ce -\frac{1}{16\pi G}\int_{\M}\d^4x\,  \left(\sqrt{|g|}\,\T(g, \Gamma) + \tilde{\Pi}\du{\alpha}{\mu\nu\rho} R\ud{\alpha}{\mu\nu\rho} + \tilde{\Xi}^{\alpha\mu\nu} Q_{\alpha\mu\nu}\right)\,.
\end{align}
Here, $\tilde{\Pi}\du{\alpha}{\mu\nu\rho}$ and $\tilde{\Xi}^{\alpha\mu\nu}$ are again tensor densities of weight $+1$, which act as Lagrange multipliers in order to reduce the general affine connection $\G{\alpha}{\mu\nu}$ to a flat and metric-compatible one. Moreover, $\T$ denotes the so-called torsion scalar and it is explicitly given by
\begin{align}\label{eq:TorsionScalar}
	\T \ce -\frac14 T_{\alpha\mu\nu} T^{\alpha\mu\nu} - \frac12 T_{\alpha\mu\nu}T^{\mu\alpha\nu} + T_\alpha T^\alpha\,.
\end{align}
The tensor $T_\alpha \ce T\ud{\lambda}{\alpha\lambda}$ denotes the only possible trace of the torsion tensor. Moreover, just as in the case of STEGR, one can prove~\cite{Heisenberg:2018, BeltranJimenez:2018, BeltranJimenez:2019} a geometric identity which relates the Ricci scalar of the Levi-Civita connection to the torsion scalar: 
 \begin{align}
 	\R = -\T - \D_\alpha T^\alpha\,. 	
 \end{align}
From this identity, one can infer that the action~\eqref{eq:TEGRAction} is equal to the Einstein-Hilbert action of GR, up to a boundary term. Thus, it follows that GR and TEGR possess the same field equations and the theories are equivalent in the sense that they possess the same solutions. The equivalence is reinforced by the fact that TEGR propagates two degrees of freedom, as has been shown in~\cite{Ferraro:2016}.

\begin{figure}[hbt!]
	\centering
	\includegraphics[width=1\columnwidth]{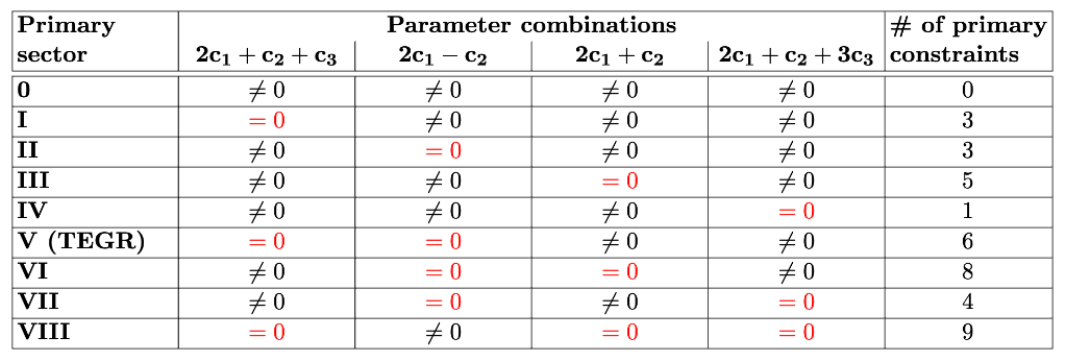}
	\caption{\protect \textit{Each primary sector (first column) is defined by the vanishing of certain parameter combinations (columns two through five). The vanishing of one or more than one of these parameter combinations corresponds to the appearance of one or more than one primary constraints. The number of independent primary constraints is listed in the last column. TEGR is contained in sector V. This table was inspired by~\cite{Blixt:2020}, but uses a slightly different nomenclature for the primary sectors.}\hspace*{\fill}}
	\label{fig:PrimarySectorsTG}
\end{figure}

The framework of metric-affine geometries with flat and metric-compatible connection can be taken as starting point for defining modified theories of gravity. Just as with the extensions of STEGR, there are two classes of theories which are commonly studied in the literature. The first class is based on an extension of the torsion scalar. Due to the anti-symmetry of the lower indices of the torsion tensor $T\ud{\alpha}{\mu\nu}$, there are only three independent contractions one can build which are quadratic in torsion. Thus, the most general quadratic torsion scalar one can construct reads
\begin{align}\label{eq:GeneralTorsionScalar}
	\T = c_1\, T_{\alpha\mu\nu} T^{\alpha\mu\nu} c_2\, T_{\alpha\mu\nu}T^{\mu\alpha\nu} + c_3\, T_\alpha T^\alpha\,,
\end{align} 
which obviously reduces to~\eqref{eq:TorsionScalar} for the parameter values $c_1 = -\frac14$, $c_2 = -\frac12$, and $c_3~=~1$. We refer to the three-parameter family of gravity theories based on the most general quadratic torsion scalar~\eqref{eq:GeneralTorsionScalar} as Teleparallel Gravity, or TG for short. The corresponding action functional is given by
\begin{align}\label{eq:TGAction}
	\S_\text{TG}[g, \Gamma; \tilde{\Pi}, \tilde{\Xi}] \ce -\frac{1}{16\pi G}\int_{\M}\d^4x\,  \left(\sqrt{|g|}\,\T(g, \Gamma; c_i) + \tilde{\Pi}\du{\alpha}{\mu\nu\rho} R\ud{\alpha}{\mu\nu\rho} + \tilde{\Xi}^{\alpha\mu\nu} Q_{\alpha\mu\nu}\right)\,,
\end{align}
where $\T(g, \Gamma; c_i)$ indicates that $\T$ is constructed from the metric, the affine-connection, and that it depends on the parameters $c_1$, $c_2$, and $c_3$. Depending on how one chooses these parameters, one can get more or less degrees of freedom compared to GR and it needs to be explored, which theories are healthy and viable modifications of gravity. A first step in this direction has been taken in~\cite{Blixt:2018, Blixt:2019}. The methodology employed in~\cite{Blixt:2018, Blixt:2019} is similar to the one used in~\cite{DAmbrosio:2020}, which led to the compartmentalization of the five-parameter family of quadratic non-metricity theories into nine sectors. That is, the authors of~\cite{Blixt:2018, Blixt:2019} determined how many primary constraints can appear from the action~\eqref{eq:TGAction} of TG depending on how one chooses the parameters $c_i$. Remarkably, it was found that TG compartmentalizes into nine sectors. This is the same number as in STG~\cite{DAmbrosio:2020}, even though TG has two parameters less than STG. A summary of the different categories of theories is given in Figure~\ref{fig:PrimarySectorsTG}.
The analysis of TG has not been extended beyond cataloging its possible primary constraints. In particular, the consistency of these constraints and the possible occurrence of secondary constraints has not yet been investigated~\cite{Blixt:2020}. As we will discuss in subsection~\ref{ssec:ObstaclesOtherTheories}, TG as well STG may suffer from the same shortcoming of the Dirac-Bergmann algorithm and completing this analysis may therefore be difficult, if not impossible. 

Finally, the non-linear extension of TEGR, which is defined by the action functional
\begin{align}\label{eq:f(T)Action}
	\S_{f(\T)}[g, \Gamma; \tilde{\Pi}, \tilde{\Xi}] \ce -\frac{1}{16\pi G}\int_{\M}\d^4x\,  \left(\sqrt{|g|}\,f(\T) + \tilde{\Pi}\du{\alpha}{\mu\nu\rho} R\ud{\alpha}{\mu\nu\rho} + \tilde{\Xi}^{\alpha\mu\nu} Q_{\alpha\mu\nu}\right)\,,
\end{align}
where $\T$ is the torsion scalar~\eqref{eq:TorsionScalar}, \textit{not} the most general scalar~\eqref{eq:GeneralTorsionScalar}, and $f$ is a function which is subjected to the sole condition that $\frac{\d f}{\d \T}\neq 0$. This condition follows from an inspection of the field equations derived from the action~\eqref{eq:f(T)Action} and the requirement of non-trivial dynamics.

The question about how many degrees of freedom propagate in $f(\T)$ gravity has received more attention than the analogous question in $f(\Q)$ gravity. However, the situation is similarly unsatisfying because it led to disputed results~\cite{Blixt:2020}. For instance, the analyses in~\cite{Li:2011, Blagojevic:2020} reached a different conclusion (five degrees of freedom) than the analysis in~\cite{Ferraro:2018} (three degrees of freedom), which may be traced back to problems in computing the Poisson brackets between constraints~\cite{Blagojevic:2020}. Furthermore, the authors of~\cite{Blagojevic:2020} also point toward the possibility of having four degrees of freedom or none at all, even if these scenarios seem unlikely.

\section{The Dirac-Bergmann Algorithm for Constrained Hamiltonian Systems}\label{sec:DiracBergmannAlgorithm}
The theory of constrained Hamiltonians was systematically developed in the 1950s by Dirac, and Bergmann and collaborators (see eg.~\cite{Dirac:1950,Bergmann:1951,DiracBook} and~\cite{HenneauxBook} and references therein for a more modern account). These developments culminated in the Dirac-Bergmann algorithm, which is used to uncover all constraints of a given Hamiltonian system. This is a necessary step to (a) check the internal consistency of a theory (eg. check whether there are ghosts), (b) define the physical (i.e., Dirac) observables of the theory, and (c) to perform the canonical quantization of the system.

Interestingly though, the algorithm was mostly developed by studying finite-dimensional system, not field theories. The transition from a finite-dimensional system to a field theory is usually regarded as a straightforward process with only minor technical complications~\cite{DiracBook, HenneauxBook, Wipf:1993} (see \cite{SundermeyerBook} for a more critical assessment of this transition).

We follow the traditional path and present the Dirac-Bergmann algorithm for finite-dimensional systems, only to then point out a little-known complication which occurs in field theories. This complication has far reaching consequences, as it has the power to invalidate the whole approach. 


\subsection{Summary of the Algorithm}\label{ssec:SummaryAlgorithm}

\begin{figure}[ht]
	\centering
	\includegraphics[keepaspectratio=true,  scale=0.6]{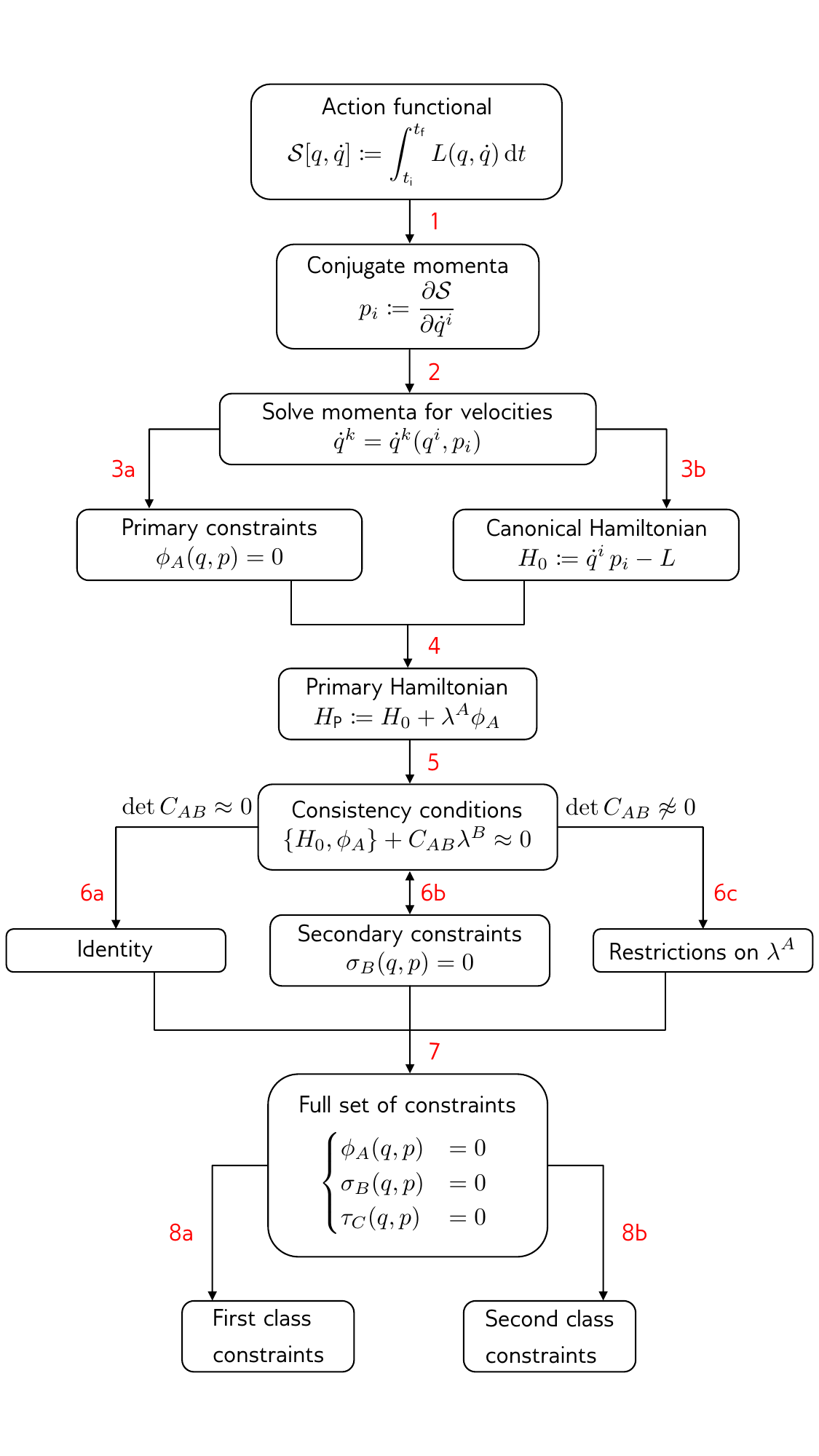}
	\caption{\protect Summary of the Dirac-Bergmann algorithm for finite-dimensional constrained Hamiltonian systems. The diagram was inspired by~\cite{Blixt:2020}.\hspace*{\fill}}
	\label{fig:DBA}
\end{figure} 

We follow roughly~\cite{Wipf:1993, Blixt:2020} in what follows and the diagram~\ref{fig:DBA}, which shows all steps of the Dirac-Bergmann algorithm, was inspired by~\cite{Blixt:2020}.

Let $\mathcal{Q}$ be a $N$-dimensional smooth manifold coordinatized by $q^{i}$ with $i\in\{1,\dots, N\}$. We refer to this manifold as \textit{configuration space}. Given $\mathcal{Q}$, we can construct the tangent bundle~$T\mathcal{Q}$ which consists of pairs $(q^{i}, \dot{q}^{i})$ and which is therefore referred to as \textit{velocity phase space}. A \textit{Lagrangian} is a function $L:T\mathcal{Q}\to \mathbb R$ which we shall assume to be at least twice smoothly differentiable, i.e., $L\in C^2(T\mathcal{Q})$. To define the theory we work with, we introduce the \textit{action functional} $\mathcal{S}:T\mathcal{Q}\to \mathbb R$ defined by
\begin{align}\label{eq:ActionFunctional}
	\mathcal{S}[q, \dot{q}] \ce \int_{t_{\textsf{i}}}^{t_{\textsf{f}}} L(q, \dot{q})\, \dd t\,.
\end{align}
The starting point for defining the Hamiltonian theory, and also step~\red{1} in diagram~\ref{fig:DBA}, is to determine the \textit{conjugate momenta} $p_i$, defined as
\begin{align}\label{eq:DefMomenta}
	p_i \ce \frac{\partial L}{\partial \dot{q}^{i}}\,.
\end{align}
The goal is to perform a Legendre transformation of the Lagrangian, which replaces the velocities $\dot{q}^{i}$ by the momenta $p_i$. However, this is only possible if the Hessian matrix
\begin{align}
	\mathcal{H}_{ij} \ce \PD{p_i}{\dot{q}^{j}} = \frac{\partial^2 L}{\partial \dot{q}^{j} \partial \dot{q}^{i}}
\end{align}
is invertible. Concretely, if $\mathcal{H}_{ij}$ has full rank\footnote{Here and in what follows we make the simplifying assumption that $\text{rank}(\mathcal{H}_{ij})$ is constant on all of $T\mathcal{Q}$.} (i.e., if $\text{rank}(\mathcal{H}_{ij}) = N$), then the equations $p_i = \partial L(q, \dot{q})/\partial \dot{q}^{i}$ can all be solved for $\dot{q}^{i}$, which means we can express all $\dot{q}^{i}$ in terms of~$q^{i}$ and~$p_i$. However, if one finds $\text{rank}(\mathcal{H}_{ij}) = R < N$, it is \textit{not} possible to replace all velocities~$\dot{q}^{i}$ by momenta $p_i$. In fact, if the Hessian has less than full rank, it means that not all conjugate momenta are independent and/or that the right hand side of~\eqref{eq:DefMomenta} contains \textit{no} velocities for some of the momenta. In either case, we can solve for only $R$ velocities,
\begin{align}\label{eq:SolvableVelocities}
	\dot{q}^{\hat{\iota}} = \mathcal{F}(q, p_{\hat{\iota}}, \dot{q}^{\bar{\iota}})\quad\text{with } \hat{\iota}\in\{1,\dots, R\} \text{ and } \bar{\iota} \in\{R+1,\dots, N\}\,.
\end{align}
We follow~\cite{Blixt:2020} and label the solvable velocities $\dot{q}^{\hat{\iota}}$ by the index $\hat{\iota}$, while the velocities one can not solve for are labeled by the index $\bar{\iota}$, i.e., $\dot{q}^{\bar{\iota}}$.

The right hand side of~\eqref{eq:DefMomenta} is in general a function $\pi_i$ of $q^{i}$, $\dot{q}^{\hat{\iota}}$, and $\dot{q}^{\bar{\iota}}$. We can thus write
\begin{align}\label{eq:SolvedMomenta}
	p_i = \PD{L}{\dot{q}^{i}} \equiv \pi_i(q^{i}, \dot{q}^{\hat{\iota}}, \dot{q}^{\bar{\iota}}) = \pi_i(q^{i}, \mathcal{F}(q, p_{\hat{\iota}}, \dot{q}^{\bar{\iota}}), \dot{q}^{\bar{\iota}}) = \pi_i(q^{i}, p_{\hat{\iota}})\,,
\end{align}
where we used equation~\eqref{eq:SolvableVelocities} and the fact that $\pi_i$ can no longer depend on $\dot{q}^{\bar{\iota}}$ after using~\eqref{eq:SolvableVelocities}, since otherwise it would be possible to solve for more velocities. Notice that for $i\in\{1,\dots, R\}$, the equations~\eqref{eq:SolvedMomenta} reduce to identities of the form $p_i \equiv p_i$. This is a direct consequence of the definition of the momenta and the fact that we could solve for~$R$ velocities. However, a different situation presents itself for the momenta with indices $i\in\{R+1, \dots, N\}$. Here we find that the right hand side gives us a relation between configuration space variables $q^{i}$ and canonical momenta $p_i$. Thus, we are forced to introduce the so-called \textit{primary constraints}
\begin{align}\label{eq:PrimaryEqs}
	\phi_A(q, p) \ce p_A - \pi_A(q, p) = 0\quad \text{for } A\in\{1, \dots, M\}\,.
\end{align}
The number $M$ of primary constraints is determined by the dimension of the configuration space and the rank of the Hessian as $M = N - R$. We stress that these $M$ relations between the configuration variables $q$ and the momenta $p$ are purely a consequence of the definition~\eqref{eq:DefMomenta}, which in turn depends on the precise form of the Lagrangian. No equations of motions were used up to this point.

However, the presence of $M$ constraints (or relations) between $q$ and $p$ implies that the dynamics of our system does \textit{not} take place in the $2N$-dimensional \textit{momentum phase space}~$\Gamma$ coordinatized by $(q^{i}, p_i)$, but rather on the submanifold $\Gamma_\textsf{P}\subseteq \Gamma$ of dimension $2N-M$. Concretely, the so-called \textit{primary constraint surface} $\Gamma_\textsf{P}$ is defined as the region of $\Gamma$ in which the equations~\eqref{eq:PrimaryEqs} are satisfied. 

Step~\red{3b} is to introduce the so-called \textit{canonical Hamiltonian}
\begin{align}
	H_0(q,p) \ce \dot{q}^{i} p_i - L(q, \dot{q})\,,
\end{align}
where a summation over the index $i\in\{1,\dots, N\}$ is implied. As one would expect, the Legendre transform guarantees that this Hamiltonian is a function of $q^{i}$ and $p_i$, i.e., it does not depend on the velocities $\dot{q}^{i}$. However, this Hamiltonian knows nothing about the primary constraints $\phi_A$ and thus does \textit{not} generate the same dynamics as the Euler-Lagrange equations derived from the functional~\eqref{eq:ActionFunctional}. To remedy this shortcoming, step~\red{4} in the Dirac-Bergmann algorithm consists in introducing $N-R$ \textit{Lagrange multipliers} $\lambda^{A}$ and to define the \textit{primary Hamiltonian}
\begin{align}
	H_{\textsf{P}}(q, p) \ce H_0 + \lambda^{A} \phi_A(q,p)\,.
\end{align}
It should be stressed that $\lambda^{A}$ are \textit{arbitrary functions of time} and their sole purpose is to enforce the primary constraints. Furthermore, one can now show that $H_\textsf{P}$ generates the same dynamics as the Euler-Lagrange equations. In particular, the time evolution of any phase space function $F:\Gamma\to \mathbb{R}$ is given by
\begin{align}\label{eq:TimeEvolution}
	\dot{F} = \{F, H_\textsf{P}\} \approx \{F, H_0\} + \lambda^{A}\{F, \phi_A\}\,,
\end{align}
where the symbol ``$\approx$'' stands for ``weak equality''. The concept of weak equality is introduced to say that the left and right hand side are equal \textit{up to terms involving arbitrary linear combinations of constraints}. In equation~\eqref{eq:TimeEvolution}, the curly brackets denote the usual Poisson bracket,
\begin{align}
	\{F, G\} \ce \PD{F}{q^{i}}\PD{G}{p_{i}} - \PD{F}{p_i}\PD{G}{q^{i}}\,,
\end{align}
and we dropped the term $\{F, \lambda^{A}\}\phi_A$, because it represents a linear combination of constraints.

For the time evolution to be consistent, i.e., for the evolution to always take place on the constraint surface $\Gamma_\textsf{P}$ and never leave it, it is necessary to impose the consistency condition
\begin{align}\label{eq:ConsistencyCondition}
	\dot{\phi}_A = \{\phi_A, H_{\textsf{P}}\} \approx \{\phi_A, H_0\} +  \{\phi_A, \phi_B\}\lambda^{B} \overset{!}{\approx} 0\,.
\end{align} 
This is step~\red{5} in Figure~\ref{fig:DBA}. It is convenient to define the $M$-dimensional vector $h_A\ce \{\phi_A, H_0\}$ and the anti-symmetric $M\times M$ matrix $C_{AB} \ce \{\phi_A, \phi_B\}$. Equation~\eqref{eq:ConsistencyCondition} can be read as system of inhomogeneous linear equations for the Lagrange multipliers $\lambda^{A}$. Depending on whether $C_{AB}$ is invertible and the internal consistency of this system of equations, there are three possible outcomes:
\begin{itemize}
	\item[(a)] Step~\red{6a}: If $\det(C_{AB}) \approx 0$, it is \textit{not} possible to solve for all Lagrange multipliers. If $\text{rank}(C) = P<M$, then there are $M-P$ vectors $u^{A}_\rho$ (with $\rho\in\{1,\dots, M-P\}$) which are (left) null vectors of $C_{AB}$. That is, these vectors satisfy
	\begin{align}
		u^{A}_\rho C_{AB} = 0\,.
	\end{align}
	Then one can show that one can solve for $P$ Lagrange multipliers if and only if 
	\begin{align}\label{eq:ConsistencyInhomogeneousSystem}
		u^{A}_\rho h_A \approx 0\,.
	\end{align}
	If this condition is \textit{not} satisfied, then the inhomogeneous system of linear equations is \textit{inconsistent} in the sense that after a Gaussian row-reduction one finds at least one equation of the form $0=1$.
	\item[(b)] Step~\red{6b}: If $\det(C_{AB})\approx 0$ and $u^{A}_\rho h_A \approx 0$, it is possible that these consistency conditions are trivially satisfied or that they lead to \textit{new constraints}, which we call \textit{secondary constraints} $\sigma_B(q,p) = 0$.
	\item[(c)] Step~\red{6c}: Finally, if $\det(C_{AB}) \not\approx 0$, equation~\eqref{eq:ConsistencyCondition} can be solved by inverting $C_{AB}$ and one finds that the Lagrange multipliers are given by
	\begin{align}
		\lambda^{A} = - C^{AB}h_B\,.
	\end{align}
\end{itemize}
We stress that the secondary constraints which arise in (b) require the validity of the equations of motion, unlike the primary constraints. If secondary constraints are encountered, their consistency under time evolution has to be checked. This can give rise to tertiary constraints and so on. In practice, the verification of consistency conditions has to be iterated, as indicated in Figure~\ref{fig:DBA} in step~\red{6b}, until no new constraints appear (or until one encounters an inconsistency).

Assuming that no inconsistencies appeared, one is now left with a full set of constraints after step~\red{7}:
\begin{align}\label{eq:SystemOfConstraint}
	\begin{cases}
		\phi_A(q,p) & = 0\\
		\sigma_B(q,p) & = 0\\
		\tau_C(q,p) & = 0\\
		&\vdots 
	\end{cases}
\end{align}
This set of constraints defines a submanifold $\Gamma_\textsf{phys}$ of the initial phase space $\Gamma$. The time evolution of physical degrees of freedom takes place entirely in $\Gamma_\textsf{phys}$. In order to be able to count the degrees of freedom, it is convenient to group the above set of constraints into two classes, as indicated in~\red{8a} and~\red{8b}: The set of \textit{first class constraints} is the maximal set consisting of all constraints $\Phi_I$ from~\eqref{eq:SystemOfConstraint} for which one has
\begin{align}
	\{\Phi_I, \Phi_J\} \approx 0 \quad\text{for all } I, J\text{ in the set of first class constraints}\,.
\end{align}
We use the abbreviation FCC for first class constraints. If a constraint is not first class, it is automatically second class (SCC). We denote the set of all second class constraints by~$\rchi_K$. One can show that the Lagrange multipliers for SCC can all be solved for. This means they are explicitly given in terms of known functions of $q^{i}$ and $p^{i}$. The same is not true for FCC, which in fact correspond to gauge transformation. This means that the Lagrange multipliers associated with FCC can be seen as the generators of such gauge transformations. It is possible to choose these Lagrange multipliers at will, without affecting physical observables. Thus, one finds that the number of degrees of freedom is given by
\begin{align}\label{eq:DOFs}
	\#\text{Degrees of freedom} = N - \# \text{FCC} - \frac12\#\text{SCC}\,.
\end{align}
Notice that FCC ``kill twice'': Each FCC removes one degree of freedom because it is a constraint and it removes a second degree of freedom because there is an arbitrarily specifiable Lagrange multiplier associated with that FCC. Also, the number of SCC must be even for the interpretation of FCC constraints (and consequently equation~\eqref{eq:DOFs}) to be valid (see~\cite{Wipf:1993} for details). This concludes the summary of the Dirac-Bergmann algorithm for particle theories. In the next subsection, we discuss a shortcoming of this algorithm which occurs in field theories.

\subsection{Field Theories: Cases in which the Algorithm fails}\label{ssec:FailureAlgorithm}
In the previous subsection we summarized the Dirac-Bergmann algorithm for systems defined over a finite dimensional configuration space. The transition to field theories is typically regarded as a task which poses only minor technical complications~\cite{DiracBook, HenneauxBook, Wipf:1993}.

Indeed, if mathematical subtleties are ignored, as can be done for most field theories of physical interest, it is straightforward to generalize steps~\red{1} through~\red{4}. However, when imposing the consistency conditions in step~\red{5} and when attempting to analyze these conditions, one can encounter serious obstacles which invalidate the whole approach.

 To give a first idea of what can go wrong in the Dirac-Bergmann analysis, we give a qualitative description of the problem and relegate an explicit example to subsection~\ref{ssec:DiracBergmannAnalysis}. First of all, the phase space variables of field theories depend on time \textit{and} space, unlike their point-particle theory relatives, where the phase space variables can only depend on time. It follows that although primary constraints cannot contain any time derivatives, they \textit{can} (but do not necessarily do) contain spatial derivatives of configuration space variables. If such spatial derivatives are present in primary constraints, this can affect the mathematical form of the consistency conditions obtained in step~\red{5}. For a field theory, the Poisson bracket is defined as
 \begin{align}
 	\{F(\vec{x}), G(\vec{y})\} \ce \int_{\Sigma_t} \dd^3z \sum_k\left(\frac{\delta F(\vec{x})}{\delta \Psi^{k}(\vec{z})} \frac{\delta G(\vec{y})}{\delta \tilde{\Pi}_k(\vec{z})} - \frac{\delta F(\vec{x})}{\delta \tilde{\Pi}_k(\vec{z})} \frac{\delta G(\vec{y})}{\delta\Psi^{k}(\vec{z})} \right)\,,
 \end{align}
 where $\Psi^{k}$ denotes the configuration space variables (i.e., the fundamental fields of the theory in question), $\tilde{\Pi}_k$ are the conjugate momentum densities associated with these fields, and $\vec{x}$, $\vec{y}$, $\vec{z}$ refer to three distinct points on the spacelike hypersurface $\Sigma_t$ (a surface of ``constant time $t$''). Similarly, the primary Hamiltonian is schematically defined as
 \begin{align}
 	H_\textsf{P} \ce \int_{\Sigma_t}\dd^3 x\, \left(\dot{\Psi}^{k}\tilde{\Pi}_k - \mathcal{L} + \lambda^{A}(t,\vec{x})\tilde{\phi}_A(\Psi, \partial\Psi, \tilde{\Pi})\right) \,,
 \end{align}
 where $\mathcal{L}$ denotes the Lagrangian density of the theory in question, $\tilde{\phi}_A(\Psi, \partial\Psi, \tilde{\Pi})$ are the constraint densities which we allow to explicitly depend on spatial derivatives of the fields~$\Psi^{k}$, and $\lambda^{A}$ are arbitrary Lagrange multipliers which depend on \textit{space and time}.
 
 In terms of the Poisson bracket and primary Hamiltonian we have just defined, the consistency condition for the primary constraints reads
 \begin{align}
 	\dot{\tilde{\phi}}_A = \{\tilde{\phi}_A, H_\textsf{P}\} = \{\tilde{\phi}_A, H_0\} + \int_{\Sigma_t}\dd^3 x\,\{\tilde{\phi}_A, \tilde{\phi}_B\}\,\lambda^{B}(t,\vec{x})\,.
 \end{align}
 Notice the crucial difference to equation~\eqref{eq:ConsistencyCondition}: The Lagrange multiplier appears inside an integral. Since $\tilde{\phi}$ depends on $\partial\Psi$, it is necessary to perform an integration by parts in order to be able to compute the variation $\delta \partial\Psi(x) / \delta\Psi(z)$ inside the Poisson bracket. However, such an integration by parts will inevitably move the spatial derivative from $\Psi$ onto $\lambda$. The consequence is that we no longer end up with an inhomogeneous system of linear equations for the Lagrange multipliers. Rather, we are now confronted with the much more difficult task of having to solve an inhomogeneous system of first order partial differential equations for the Lagrange multipliers. Recall that the original identification of secondary and tertiary constraints in the Dirac-Bergmann algorithm relies on the fact that one has to analyze a system of linear algebraic equations. Therefore, in the case we just considered, we can no longer proceed with the steps~\red{6} through~\red{8}.
 
Of course, we do not actually need to solve the first order system of PDEs. We only need to determine whether or not a unique solution exists. This is formally possible, but conceptual problems arise nevertheless: In order to obtain a unique solution we also need to impose initial value conditions. What do these conditions physically correspond to? Do different choices of initial conditions manifest themselves physically? That is, do different initial conditions represent physically inequivalent situations? 

The situation is even worse when it can be shown that no unique solution exists and that additional constraints arise. In addition to the conceptual problems alluded to above, one faces also considerable technical difficulties in identifying secondary constraints and checking the consistency of the combined set of primary and secondary constraints.

It is conceivable that there are theories which are simple enough such that these problems can be solved in a consistent way. However, in general, the Dirac-Bergmann algorithm breaks down at this point. Ignoring this issue likely leads to wrong conclusions.

In the next subsection we show explicitly how this issue invalidates the Dirac-Bergmann analysis of $f(\Q)$ gravity.

\subsection{Dirac-Bergmann Analysis of $f(\Q)$ Gravity in the Coincident Gauge}\label{ssec:DiracBergmannAnalysis}
As evident from Figure~\ref{fig:DBA}, the first step in the Dirac-Bergmann algorithm consists of determining the conjugate momenta starting from the action functional. To that end, one first needs to define the time derivative of configuration space variables, from which the momenta can then be constructed. In any covariant theory, this means that spacetime~$(\mathcal M, g_{\mu\nu})$ needs to be arbitrarily foliated by spacelike leaves $\Sigma_t$. Concretely, we assume that~$\mathcal M$ has the topology $\mathcal M\cong \mathbb R\times \Sigma$, where $\Sigma$ is a three-dimensional manifold of arbitrary topology representing ``space'', while $\mathbb R$ represents ``time''.

Any coordinate system $x^\mu$ on $\mathcal M$ can then be split as $(t, x^{a})$, where $t$ is the ``time'' coordinate and $x^{a}$ are the ``spatial'' coordinates\footnote{Our convention is that Latin indices such as $a,b$ range from $1$ to $3$, while Greek letters denote spacetime indices and correspondingly range from $0$ to $3$.}. The leaves $\Sigma_t$ simply correspond to~$t= const.$  hypersurfaces. Moreover, it is convenient to write the metric in a form adapted to the arbitrarily chosen foliation of spacetime. This so-called Arnowitt-Deser-Misner (ADM) form of the metric is given by
\begin{align}
	g_{\mu\nu} = 
	\begin{pmatrix}
		-N^2 + h_{ab} N^{a} N^{b} & h_{ab}N^{b} \\
		h_{ab}N^{b} & h_{ab}
	\end{pmatrix}\,,
\end{align}
where $N$ denotes the lapse function\footnote{We recall that $N\neq 0$ is required in order to obtain a well-defined foliation of spacetime. Without loss of generality, we assume $N>0$ everywhere.}, $N^{a}$ is the shift vector field, while $h_{ab}$ denotes the intrinsic spatial metric on the spacelike leaves of our foliation.

In what follows we work with the ADM variables $\{N, N^{a}, h_{ab}\}$ and we employ the coincident gauge, i.e., we exploit the fact that we can choose $\G{\alpha}{\mu\nu} = 0$ everywhere\footnote{In subsections~\ref{sec:TeleparallelTheories} and~\ref{ssec:KineticMatrixApproach} we give a more detailed account of why this is always possible.}. These choices allow us to write the $f(\Q)$ action functional as
\begin{align}\label{eq:f(Q)ActionInADMVariables}
	\S[N, N^{a}, h_{ab}, \phi] \ce \int_\M \d^4 x\, \sqrt{h}\, N\left(f(s) -\phi (s-\Q)\right)\,,
\end{align}
where we used the easy to prove fact that
\begin{align}
	\sqrt{|g|} = N\, \sqrt{h}\,,
\end{align}
and where we introduced an auxiliary field $\phi$ in order to ``pull out'' the non-metricity scalar~$\Q$ from the function $f$. This has the advantage that we can use results obtained in~\cite{DAmbrosio:2020} regarding variations of $\Q$ with respect to the velocities $\dot N$, $\dot N^{a}$, and $\dot{h}_{ab}$, which leads to particularly simple forms of the conjugate momenta. We stress that \textit{(a)} the action~\eqref{eq:f(Q)ActionInADMVariables} is equivalent to the original $f(\Q)$ action~\eqref{eq:f(Q)Action} written in ADM variables and in coincident gauge and \textit{(b)} that we need to add a boundary term in order to use the ADM form of~\cite{DAmbrosio:2020}, with the crucial difference that $\phi$ will inevitably enlarge the phase space and derivatives will act on it. The addition of this boundary term has no effect on the counting of degrees of freedom since it does not modify the field equations. It will just simplify the expressions. Furthermore, the construction has been explicitly carried out in~\cite{Hu:2022}, which thus prompts us to refrain from going through the individual steps. \newline

\noindent\textbf{Step \red{1} (conjugate momenta):}\\
After having introduced the necessary framework for a Hamiltonian analysis, defined variables adapted to our purpose, and moulded the action functional into a more convenient form, we can finally execute the first step of the Dirac-Bergmann algorithm: Determining the momenta conjugate to $N$, $N^{a}$, $h_{ab}$, and $\phi$. Notice that these momenta are necessarily tensor densities, since the Lagrangian from which they are derived is a scalar density. We place a tilde on top of the momenta to emphasize this fact and we find that they have the form one would expect from the results obtained in~\cite{DAmbrosio:2020}, namely
\begin{align}\label{eq:MomentumDensities}
	\tilde{\pi} &\ce \varD{\S}{\dot N} = 0\, ,& \tilde{\pi}^{ab} &\ce \varD{\S}{\dot{h}_{ab}} = \sqrt{h}\,\phi\, K_{cd}\left(h^{ac} h^{bd} - h^{ab} h^{cd}\right)\notag\\
	\tilde{\pi}_a &\ce \varD{\S}{\dot{N}^{a}} = \frac{\sqrt{h}}{N}\partial_a \phi\, , & 	\tilde{\pi}_\phi &\ce \varD{\S}{\dot\phi} = -\frac{\sqrt{h}}{N}\partial_a N^{a}\,.
\end{align}
Here, $K_{ab}$ denotes the extrinsic curvature, which is defined as
\begin{align}
	K_{ab} \ce \frac{1}{2N}\left(\mathfrak D_{(a} N_{b)} - \dot{h}_{ab} \right)\,,
\end{align}
and where $\mathfrak D_a$ is the unique torsion-free covariant derivative operator compatible with $h_{ab}$.\newline

\noindent\textbf{Step \red{2} (solve for velocities):}\\
Observe that only the momentum density $\tilde{\pi}^{ab}$ contains a velocity, namely $\dot{h}_{ab}$, which is ``hidden'' inside the extrinsic curvature tensor $K_{ab}$. The velocities $\dot{N}$, $\dot{N}^{a}$, and $\dot{\phi}$ do not appear in any of the momenta. Therefore, we can only solve for $\dot{h}_{ab}$, for which we obtain
\begin{align}
	\dot{h}_{ab} = \mathfrak D_{(a}N_{b)} - \frac{2N}{\sqrt{h}\, \phi}\left(\tilde{\pi}_{ab} - \frac12 h_{ab} \tilde{\pi}\ud{c}{c}\right)\,.
\end{align}\newline

\noindent\textbf{Step \red{3a} (primary constraints):}\\
Since the momentum densities~\eqref{eq:MomentumDensities} do not depend on the velocities $\dot{N}$, $\dot{N}^{a}$, and $\dot{\phi}$, we obtain a total of $1+3+1=5$ primary constraints, which we denote by
\begin{align}\label{eq:PrimaryConstraints}
	\tilde{C}_0 &\ce \tilde{\pi} \approx 0\,, & \tilde{\C}_a &\ce \tilde{\pi}_a - \frac{\sqrt{h}}{N} \partial_a \phi \approx 0\,, & \tilde{C}_\phi &\ce \tilde{\pi}_\phi + \frac{\sqrt{h}}{N}\partial_a N^{a} \approx 0\,.
\end{align}\newline

\noindent\textbf{Step \red{3b} (canonical Hamiltonian):}
To define the canonical Hamiltonian $H_0$, it is convenient to first introduce its density:
\begin{align}
	\mathcal{H}_0 &\ce \dot{N}\,\tilde{\pi} + \dot{N}^{a}\, \tilde{\pi}_a + \dot\phi\, \tilde{\pi}_\phi + \dot{h}_{ab} \,\tilde{\pi}^{ab} - \mathcal{L}\,.
\end{align}
The canonical Hamiltonian is then defined as the integral of this density over one of the leaves $\Sigma_t$ in the foliation of spacetime, i.e.,
\begin{align}\label{eq:CanonicalHamiltonian}
	H_0(\Sigma_t) &\ce \int_{\Sigma_t} \dd^3 x\, \mathcal{H}_0\notag\\
	&\approx \int_{\Sigma_t}\dd^3 x\bigg[\frac{N}{\sqrt{h}\,\phi}\left(\frac12 \tilde{\pi}\ud{c}{c}\tilde{\pi}\ud{d}{d} - \tilde{\pi}_{ab}\tilde{\pi}^{ab}\right)\notag\\
	&\phantom{\approx\int_{\Sigma_t}\dd^3} + \sqrt{h} N \left(\prescript{(3)}{}{}R\,\phi + \left\{\prescript{(3)}{}{}\bar{Q}^{a} - \prescript{(3)}{}{}Q^{a}\right\}\mathfrak D_{a} \phi + s\,\phi - f(s)\right) - 2 N_a \mathfrak D_b\tilde{\pi}^{ab}\bigg]\,.
\end{align}
The ``$\approx$'' means that we used the primary constraints~\eqref{eq:PrimaryConstraints} to bring the integrand in the above form. Moreover, we performed a partial integration and dropped the boundary term, as it will not affect the counting of degrees of freedom. We remark that if one chooses $f(\Q) = \Q$, which then also implies $\phi=1$, equation~\eqref{eq:CanonicalHamiltonian} reduces to the canonical Hamiltonian of Coincident GR which was derived in~\cite{DAmbrosio:2020}. This is the same canonical Hamiltonian as in GR, as one would expect.\newline

\noindent\textbf{Step \red{4} (primary Hamiltonian):}\\
Given that we now know all primary constraints and the canonical Hamiltonian, we can turn to the task of defining the primary Hamiltonian. To do so, we need to introduce the Lagrange multipliers $\lambda^0$, $\lambda^{a}$, and $\lambda^\phi$. The primary Hamiltonian can then be written as
\begin{align}
	H_\textsf{P}(\Sigma_t) \ce H_0(\Sigma_t) + \int_{\Sigma_t}\dd^3 x\left(\lambda^{0} \tilde{C}_0 + \lambda^{a} \tilde{C}_a + \lambda^{\phi}\tilde{C}_\phi\right)\,.
\end{align}\newline

\noindent\textbf{Step \red{5} (consistency conditions):}\\
Before checking the consistency conditions, we wish to highlight a peculiarity which arises in the context of field theories. Namely, that the Poisson bracket between two constraint densities produces a $\delta$-function or even spatial derivatives of a $\delta$-function. Expressions involving $\delta$-functions which do \textit{not} occur inside an integral have a rather formal character and they can easily lead to confusion, as we will discuss further below. 

A simple way to deal with this issue is to introduce so called \textit{smeared constraints}. Concretely, one integrates a constraint density against a test function $F$, i.e., one defines
\begin{align}\label{eq:SmearedConstraint}
	C_I[F] &\ce \int_{\Sigma_t} \d^3 x\, \tilde{C}_I(x)\, F(x)\, && \text{for }  I \in\{0, a, \phi\}
\end{align}
and for any test function $F$. When computing Poisson brackets between constraint densities, one encounters a product of two $\delta$-functions, but because of the presence of the two smearing integrals, these $\delta$-functions can be ``integrated out'', and one is left with mathematically well-defined expressions. Applying this (standard) strategy to our constraints results in the following Poisson brackets:
\begin{align}\label{eq:PBBrackets}
	\{C_0[F_1], C_a[F_2]\} &= -\int_{\Sigma_t}\d^3 z\, \frac{\sqrt{h}}{N^2}\partial_a\phi\, F_1\, F_2 \approx -\int_{\Sigma_t}\dd^3 z\, \frac{\tilde{\pi}_a}{N}\,F_1\,F_2 \notag\\
	\{C_0[F_1], C_\phi[F_2]\} &= \int_{\Sigma_t}\d^3 z\, \frac{\sqrt{h}}{N^2}\partial_a N^{a}\,F_1\, F_2 \approx - \int_{\Sigma_t} \dd^3 z\, \frac{\tilde{\pi}_\phi}{N}\, F_1\, F_2 \notag\\
	\{C_a[F_1], C_\phi[F_2]\} &= \int_{\Sigma_t}\d^3 z\left[\partial_a \left(\frac{\sqrt{h}}{N} F_1\right) F_2 + \partial_a\left(\frac{\sqrt{h}}{N} F_2\right) F_1\right]\,.
\end{align}
The integral which appears on the right hand side stems from the integral in the definition of the Poisson bracket and has nothing to do with the smearing we introduced above. 

The consistency conditions we have to check are
\begin{align}
	\{\tilde{C}_I, \Hp\} &= \underbrace{\{\tilde{C}_I, H_0\}}_{\ec \tilde{b}_I} + \int_{\Sigma_t}\d^3 z\, \{\tilde{C}_I, \tilde{C}_J\} \lambda^{J} \overset{!}{\approx} 0 &\text{for } I,J\in\{0, a, \phi\}\,,
\end{align}
and our task is to compute the quantities $\tilde{b}_I$ and the integral over the Poisson bracket. For the vector densities $\tilde{b}_I$ we find the following results:
\begin{align}
	\tilde{b}_0 &\ce \{\tilde{C}_0, H_0\} = \tilde{S}_\phi + \sqrt{h}\, N\left(f(s) - s\,\phi - \left[^{(3)}\bar{Q}^a - ^{(3)}Q^a\right]\partial_a\phi \right)\notag\\
	\tilde{b}_a &\ce \{\tilde{C}_a, H_0\} \approx \tilde{V}_a + \frac{N}{\sqrt{h}}\left(\tilde{\pi}_\phi - \frac{1}{2\phi} \tilde{\pi}\ud{b}{b}\right)\tilde{\pi}_a \notag\\
	\tilde{b}_\phi &\ce \{\tilde{C}_\phi, H_0\} \approx N\frac{\d}{\d\phi}\tilde{S}_\phi + \partial_a\left(\sqrt{h}\, N\left[^{(3)}\bar{Q}^{a} - ^{(3)}Q^{a}\right]\right) - \sqrt{h}\, N\, s \notag\\
	&\phantom{\{\tilde{C}_\phi, H_0\} \approx}\qquad + \frac{N}{\sqrt{h}} \left(\tilde{\pi}_\phi - \frac{1}{2\phi}\tilde{\pi}\ud{b}{b}\right)\tilde{\pi}_\phi\,,
\end{align}
where we have introduced
\begin{align}
	^{(3)}Q^{a} &\ce h^{ij}h^{ab}Q_{bij} \notag\\
	^{(3)}\bar{Q}^{a} &\ce h^{ij}h^{ab}Q_{ibj}
\end{align}
as well as
\begin{align}
	\tilde{\mathcal{S}}_\phi &\ce \frac{1}{\sqrt{h}\,\phi}\left(\tilde{\pi}_{ab}\tilde{\pi}^{ab} - \frac12 \tilde{\pi}\ud{a}{a}\tilde{\pi}\ud{b}{b}\right) - \sqrt{h}\,\phi\, ^{(3)}\mathcal{R} \notag\\
	\tilde{\mathcal{V}}_a &\ce 2\mathfrak{D}_b\tilde{\pi}\ud{b}{a}\,.
\end{align}
Notice that $\tilde{\mathcal{V}}_a$ is precisely the expression for the vector density constraint of standard GR, while $\tilde{\mathcal{S}}_\phi$ reduces to GR's scalar density constraint for $\phi=1$.

The integral over the Poisson bracket is easy to compute using the brackets~\eqref{eq:PBBrackets}. In fact, using definition~\eqref{eq:SmearedConstraint}, one can write this integral as
\begin{align}
	\int_{\Sigma_t}\dd^3 z\,\{\tilde{C}_I, \tilde{C}_J\} \lambda^{J} = \{\tilde{C}_I, C_J[\lambda^{J}]\} = \{C_I[\delta], C_J[\lambda^{J}]\}\,,
\end{align}
where $C_I[\delta]$ means we choose a $\delta$-function to smear $\tilde{C}_I$ and one immediately sees that $C_I[\delta] = \tilde{C}_I$.
Using~\eqref{eq:PBBrackets} with $F_1 = \delta$ and $F_2 = \lambda^{J}$, we now find the following expressions for the Poisson brackets:
\begin{align}
	\{C_0[\delta],  C_J[\lambda^J]\} &= -\frac{1}{N}\tilde{\pi}_a\lambda^{a} -\frac{1}{N}\tilde{\pi}_\phi \lambda^\phi  && \text{(sum over $a$)} \notag\\
	\{C_a[\delta],  C_J[\lambda^J]\} &= \frac{1}{N}\tilde{\pi}_a \lambda^0 + 2 \partial_a\left(\frac{\sqrt{h}}{N}\right)\lambda^\phi + \frac{\sqrt{h}}{N} \partial_a \lambda^\phi && \text{(for $a\in\{1,2,3\}$)} \notag\\
	\{C_\phi[\delta],  C_J[\lambda^J]\} &= \frac{1}{N}\tilde{\pi}_\phi \lambda^0 - 2\partial_a\left(\frac{\sqrt{h}}{N}\right)\lambda^{a} - \frac{\sqrt{h}}{N} \partial_a \lambda^{a} && \text{(sum over $a$)}\,.
\end{align}
We make the important observation that these expressions contain derivatives of the Lagrange multipliers. As explained in the previous subsection, the Dirac-Bergmann algorithm tacitly assumes that the consistency conditions lead to a system of \textit{linear} equations for the Lagrange multipliers, which can be analyzed using standard tools of linear algebra. 

Here, however, we end up with a first order system of partial differential equations (PDEs). We now explicitly show that this prevents us from making any progress in the Dirac-Bergmann analysis. First of all, the consistency conditions can now be written concisely as
\begin{align}
	\left\{\begin{array}{rll}
		-\frac{1}{N}\tilde{\pi}_a\lambda^{a} -\frac{1}{N}\tilde{\pi}_\phi \lambda^\phi + \tilde{b}_0 & \overset{!}{\approx} 0 & \qquad\text{(i)} \\
		\frac{1}{N}\tilde{\pi}_a \lambda^0 + 2 \partial_a\left(\frac{\sqrt{h}}{N}\right)\lambda^\phi + \frac{\sqrt{h}}{N} \partial_a \lambda^\phi + \tilde{b}_a &\overset{!}{\approx} 0 & \qquad\text{(ii)} \\
		\frac{1}{N}\tilde{\pi}_\phi \lambda^0 - 2\partial_a\left(\frac{\sqrt{h}}{N}\right)\lambda^{a} - \frac{\sqrt{h}}{N} \partial_a \lambda^{a} + \tilde{b}_\phi &\overset{!}{\approx} 0 & \qquad\text{(iii)}
	\end{array}\right.\,.
\end{align}
Equation (i) is purely algebraic and it can be used to eliminate $\lambda^\phi$. This results in
\begin{align}\label{eq:LambdaPhi}
	\lambda^\phi = \frac{N}{\tilde{\pi}_\phi}\tilde{b}_0 -\frac{\tilde{\pi}_a}{\tilde{\pi}_\phi}\lambda^{a}\,.
\end{align}
 After plugging this expression for $\lambda^\phi$ into equations (ii) and (iii), we are left with three PDEs for the three functions $\lambda^{a}$ and an equation for the function $\lambda^{0}$. Equation (iii) can be solved for $\lambda^{0}$, giving us
 \begin{align}\label{eq:LambdaZero}
 	\lambda^{0} = \frac{2N}{\tilde{\pi}_\phi}\partial_a\left(\frac{\sqrt{h}}{N}\right)\lambda^{a} + \frac{\sqrt{h}}{\tilde{\pi}_\phi} \partial_a \lambda^{a} - \frac{N}{\tilde{\pi}_\phi}\tilde{b}_\phi\,.
 \end{align}
We recall that in the standard Dirac-Bergmann algorithm, we are mostly interested in determining whether the system of linear equations for the Lagrange multipliers is solvable or not (see steps~\red{6a},~\red{6b}, and~\red{6c}). In our case, we can ask the same question: Is the first order system of PDEs solvable or not?

This question can be answered using standard methods from the mathematical theory of partial differential equations, which is summarized in Appendix~\ref{app:PDEs}. Here, we simply remark that the PDE (ii), after having plugged in~\eqref{eq:LambdaPhi} and~\eqref{eq:LambdaZero}, is of the form
\begin{align}\label{eq:SchematicFormPDEs}
	\tilde{\pi}_a \partial_b \lambda^{b} - \tilde{\pi}_b \partial_a \lambda^{b}  + \text{lower order terms} \overset{!}{\approx} 0\,.
\end{align}
By introducing the matrices 
\begin{align}\label{eq:MMatrices}
	M^{(1)} &\ce
	\begin{pmatrix}
		\tilde{\pi}_3 & 0 & 0 \\
		\tilde{\pi}_2 & 0 & 0 \\
		0 & -\tilde{\pi}_2 & -\tilde{\pi}_3
	\end{pmatrix}, &
	M^{(2)} &\ce 
	\begin{pmatrix}
		0 & \tilde{\pi}_3 & 0 \\
		-\tilde{\pi}_1 & 0 & -\tilde{\pi}_3 \\
		0 & \tilde{\pi}_1 & 0
	\end{pmatrix}, &
	M^{(3)} &\ce 
	\begin{pmatrix}
		-\tilde{\pi}_1 & -\tilde{\pi}_2 & 0 \\
		0 & 0 & \tilde{\pi}_2 \\
		0 & 0 & \tilde{\pi}_1
	\end{pmatrix}\,,
\end{align}
equation~\eqref{eq:SchematicFormPDEs} can be re-written as
\begin{align}\label{eq:MatrixForm2}
	\sum_{i=1}^{3} M^{(i)}\partial_i\vec{\lambda} + \text{lower order terms}\overset{!}{\approx} 0\,.
\end{align}
The question whether this system admits a unique solution for $\lambda^{a}$ can now be answered by checking whether the matrix $\sum_{i=1}^{3} M^{(i)}n_i$, where $n_i$ are the components of the normal vector to the initial value surface, has a non-vanishing determinant. However, one trivially finds
\begin{align}\label{eq:SolvabilityCriterion}
	\det\left(\sum_{i=1}^{3} M^{(i)}n_i\right) = 0\,,
\end{align}
for any nowhere vanishing vector field $\vec{n}$. A direct computation further reveals that the above matrix has rank two. Two consequences follow from this fact:
\begin{itemize}
	\item[(a)] Since the above matrix has rank two, it is possible to bring the PDE into such a form that only two equations contain derivatives of $\lambda^{a}$, while the third equation will be purely algebraic. In other words, one finds a new constraint. (This is the analogue of equation~\eqref{eq:ConsistencyInhomogeneousSystem} in the Dirac-Bergmann algorithm)
	\item[(b)] Because there are only two \textit{algebraically independent} differential equations in the system, it is only possible to determine two of the functions $\lambda^{a}$, while the third one has to be specified by hand. This could represent a gauge freedom.
\end{itemize}
However, the interpretation of this freedom is not straightforward. Moreover, should the freedom to prescribe initial data for $\lambda^{a}$ on, say, a $z=const.$ surface also be interpreted as gauge freedom? What are the consequences of having this freedom? In what way does this affect the counting of degrees of freedom?

These questions are far from trivial and we will not attempt to answer them here. Rather, we point out that the Dirac-Bergmann procedure breaks down at this point. The reason for the breakdown is the presence of spatial derivatives of configuration space variables in the primary constraints, which lead to PDEs, rather than linear equations, for the Lagrange multipliers. The silent assumption of the Dirac-Bergmann algorithm that one always obtains linear equations for the Lagrange multipliers is therefore broken.

Before concluding this subsection, we point out where the analysis of~\cite{Hu:2022} went wrong. As alluded to above, a peculiarity of field theories is that the Poisson bracket of two constraints produces a $\delta$-function or even spatial derivatives of $\delta$-functions. This is precisely what happened in~\cite{Hu:2022} (see equation~(43) in~\cite{Hu:2022}). The consistency conditions found in~\cite{Hu:2022} are then only \textit{apparently} linear equations for the Lagrange multipliers. In fact, the matrix $C_{n'n}$ in equation~(49) of~\cite{Hu:2022} contains $\delta$-functions and spatial derivatives thereof, while the vector $\{\phi_{n'},\mathcal{H}_0\}$ in equation~(48) only contains a $\delta$-function. We recall that $\delta$-functions acquire a well-defined mathematical meaning when integrated over. If the \textit{apparently} linear equation~(48) of~\cite{Hu:2022} is integrated over a $\Sigma_t$ surface, the $\delta$-functions and their derivatives can be ``integrated out'', leaving behind a system of first oder PDEs, rather than linear equations for the $\lambda$'s.

The argument presented here casts considerable doubts on the validity of the Dirac-Bergmann algorithm in the case of $f(\Q)$ gravity and the results obtained in~\cite{Hu:2022}. In the next subsection, we show that (almost) every teleparallel theory of gravity is affected by the same problem and in section~\ref{sec:UpperBound} we show explicitly, using a completely independent method, that the number of degrees of freedom in $f(\Q)$ gravity is at most seven.

\subsection{Potential Obstacles in other Teleparallel Theories of Gravity}\label{ssec:ObstaclesOtherTheories}
In the case of $f(\Q)$ gravity, we have seen how the presence of spatial derivatives in constraint densities can invalidate the whole Dirac-Bergmann approach. Here, we show that other teleparallel theories of gravity are potentially afflicted by the same problem. Concretely, we consider the five-parameter family of theories described in~\ref{ssec:FiveParameterFamily}, as well as $f(\T)$ gravity and the three-parameter family of torsion theories mentioned in~\ref{ssec:TorsionTheories}.

Starting with the most general non-metricity scalar~\eqref{eq:GeneralNonMetricityScalar}, one finds, after choosing the coincident gauge and applying a subsequent ADM decomposition of the metric, that the momentum densities conjugate to lapse, shift, and intrinsic metric can be written as
\begin{align}
	\tilde{\pi} &=  \frac{2\sqrt{h}}{N}\,n_\alpha n_\mu n_\nu P^{\alpha\mu\nu}\notag\\
	\tilde{\pi}_a &=  \frac{2\sqrt{h}}{N}\, \left(-\frac{1}{N}n_\alpha n_\mu n_\nu P^{\alpha\mu\nu} h_{ac} N^{c} + n_{\alpha} n_{\mu} P^{\alpha\mu c} h_{ac}\right)\notag\\
	\tilde{\pi}^{ab} &= \frac{\sqrt{h}}{N}\left(-\frac{1}{N}n_\alpha n_\mu n_\nu P^{\alpha\mu\nu}N^{a} N^{b} + 2 n_\alpha n_\mu P^{\alpha\mu (a}N^{b)} - Nn_\alpha P^{\alpha ab} \right)\,,
\end{align}
where $n_{\mu} = (-N, \vec{0})$ is the unit timelike normal $1$-form to $\Sigma_t$ and $P\ud{\alpha}{\mu\nu}$ is the so-called non-metricity conjugate 
\begin{align}\label{eq:NonMetricityConjugate}
	P\ud{\alpha}{\mu\nu} &\ce \frac12\PD{\Q}{Q\du{\alpha}{\mu\nu}}\notag\\
	&= c_1 Q\ud{\alpha}{\mu\nu} + c_2 Q\dud{(\mu}{\alpha}{\nu)} + c_3 g_{\mu\nu}Q^\alpha + c_4 \delta\ud{\alpha}{(\mu}\bar{Q}_{\nu)} + \frac12c_5 \left(g_{\mu\nu}\bar{Q}^\alpha + \delta\ud{\alpha}{(\mu}Q_{\nu)}\right).
\end{align}
As shown in~\cite{DAmbrosio:2020b}, these momenta can be written more explicitly as
\begin{align}
	\tilde{\pi} &= \frac{\sqrt{h}}{N^4}\bigg[2\tilde c \left( Q_{\red{000}} - N^{a} Q_{a\red{00}} - 2N^{a} Q_{\red{00}a} + 2 N^{a} N^{b} Q_{ab\red{0}} + N^{a} N^{b} Q_{\red{0} ab} - N^{a} N^{b} N^{c} Q_{abc}\right)\notag\\
	&\phantom{\frac{\sqrt{h}}{N}2c}\ \ + c_{35} \left(N^2 N^{a} Q_a - N^2 Q_{\red{0}}\right) + c_{45}\left(N^2 N^{a} \bar Q_a - N^2  \bar Q_{\red{0}}\right)\bigg]\notag\\
	\tilde{\pi}_a &= \frac{\sqrt{h}}{2N^3}\bigg[2\hat c\left(Q_{\red{00}a}-N^{b}Q_{ba\red{0}} - N^{b} Q_{\red{0} ab} + N^{b} N^{c} Q_{bac}\right) \notag\\
	&\phantom{\frac{\sqrt{h}}{2N^3}2c}\ \ + c_{25}\left(Q_{a\red{00}} -2 N^{b} Q_{ab\red{0}} + N^{b} N^{c} Q_{abc}\right) - 2 c_4 N^2 \bar Q_a - c_5 N^2 Q_a\bigg]\notag\\
	\tilde{\pi}^{ab} &= \frac{\sqrt{h}}{2 N^3}\bigg[ c_{35} \left(h^{ab} Q_{\red{000}} - h^{ab} N^{c} \left(Q_{c\red{00}} + 2 Q_{\red{00} c} - N^{d} \left\{ 2 Q_{cd\red{0}} + Q_{\red{0} cd} - N^{e}Q_{cde}\right\}\right)\right)\notag\\
	&\phantom{\frac{\sqrt{h}}{2 N^3} c_{35}}\ \ + N^2 N^{c} h^{ad} h^{be}  \left(2 c_1 Q_{cde} + c_2 Q_{dce} + c_2 Q_{ecd}\right) + N^2 N^{c} h^{ab} h^{de} \left(2c_3 Q_{cde} + c_5 Q_{dce}\right)\notag\\
	&\phantom{\frac{\sqrt{h}}{2 N^3} c_{35}}\ \ - N^2 h^{ab} h^{cd}\left(2 c_3 Q_{\red{0}cd} + c_5 Q_{cd\red{0}}\right) - 2 N^2 h^{ac} h^{bd} \left( c_1 Q_{\red{0} cd} + c_2 Q_{(cd)\red{0}}\right)\bigg]\,,
\end{align}
where we used the shorthand notations
\begin{align}
	\tilde{c} &\ce c_1 + c_2 + c_3 + c_4 + c_5\notag\\
	\hat{c} &\ce 2c_1 + c_2 + c_4 \notag\\
	c_{i5} &\ce 2c_i + c_5  & \text{ for } i\in\{2,3,4\}\,.
\end{align}
Notice that only non-metricity tensor components of the form $Q_{\red{0}\mu\nu}$ and $Q_{\red{0}}$ contain time derivatives. All other non-metricity tensor components contain spatial derivatives. The nine sectors defined in Figure~\ref{fig:PrimarySectors} correspond to choices of $c_i$ such that the time derivatives disappear from the momenta\footnote{With the exception of sector $0$, which has no constraints at all and for which the problem discussed here does definitely not arise. However, sector $0$ is pathological, as also discussed in~\cite{DAmbrosio:2020b}.}. However, none of these choices leads to a complete absence of spatial derivatives. Therefore, each primary sector described in Figure~\ref{fig:PrimarySectors} contains constraints with spatial derivatives of lapse, shift, and/or intrinsic metric. It follows that potentially every single theory in this five-parameter family of Lagrangians is affected by the same problem we encountered in $f(\Q)$ gravity, namely that the Dirac-Bergmann algorithm cannot be carried through to completion.

Notice that this is even the case for STEGR, which is contained in sector V. The momenta conjugate to lapse and shift in the case of STEGR are non-trivial and involve spatial derivatives. In~\cite{DAmbrosio:2020}, the Hamiltonian analysis could only be carried out after the STEGR Lagrangian had been modified by partial integrations, which is the same as adding boundary terms. The effect was such that the momenta conjugate to lapse and shift trivialised. However, this strategy is likely not always successful. The authors of~\cite{Hu:2022} attempted the same kind of simplifications in the case of $f(\Q)$ gravity. While they managed to trivialise the momentum conjugate to $N$, the other momenta retained a non-trivial form and they feature spatial derivatives. 

Next, we turn our attention to the three parameter family of torsion theories summarized in Figure~\ref{fig:PrimarySectorsTG}. Explicit expressions for the primary constraints\footnote{We refrain from showing these expressions here, since~\cite{Blixt:2018, Blixt:2019, Blixt:2020} use the tetrad formalism. This would require us to introduce a lot of new terminology and notation.} can be found in~\cite{Blixt:2018, Blixt:2019, Blixt:2020}. It can be verified that all primary sectors, with the exception of the pathological sector $0$, harbour primary constraints with spatial derivatives of configuration space variables. Thus, for all of these theories it is potentially not possible to carry out the Dirac-Bergmann analysis to its completion.

The situation for $f(\T)$ gravity is similar. Interestingly, the authors of~\cite{Ferraro:2018} fell victim to the same error as the authors of~\cite{Hu:2022}: When determining the consistency conditions, they overlooked the presence of spatial derivatives of $\delta$-functions and arrived at \textit{seemingly} linear equations for the Lagrange multipliers. An analysis of this erroneous linear system led them to conclude that there are three propagating degrees of freedom in $f(\T)$ gravity. 

The error was recognized in~\cite{Blagojevic:2020}, but it could not be corrected. The authors of~\cite{Blagojevic:2020} expressed puzzlement at finding a system of linear PDEs for the Lagrange multipliers and discussed three possible scenarios, giving them different numbers of degrees of freedom for~$f(\T)$. In any case, it is clear that $f(\T)$ suffers from exactly the same problem as $f(\Q)$ and that the question about the number of physical degrees of freedom remains unsettled in both theories.

Before concluding this subsection, we remark that even though it is true that \textit{all} teleparallel theories of gravity discussed here contain primary constraints with spatial derivatives, we do not claim that every single one of them necessarily leads to first order partial differential equations for its Lagrange multipliers. We believe that it is \textit{likely} that this happens in every case, but the presence of spatial derivatives in the constraints is only a \textit{necessary} and \textit{not} a sufficient condition for this to happen. In fact, there are two important examples where constraints with spatial derivatives do appear without leading to problems in the Dirac-Bergmann algorithm: Electromagnetism and GR \`{a} la Einstein. 

In electromagnetism one finds that the momentum $\Pi^{\red{0}}$ conjugate to $A_{\red{0}}$ vanishes, which thus constitutes a primary constraint. Clearly, this constraint is unproblematic as it does not contain any derivatives. However, running through the Dirac-Bergmann algorithm up to Step~\red{5}, one finds that this constraint generates the secondary constraint
\begin{align}
	\partial_a \Pi^{a} \approx 0\,,
\end{align}
where $\Pi^{a}$ is the momentum conjugate to $A_a$. This is the well-known Gauss constraint and it clearly contains a spatial derivative. The reason this does not create any problems is because the Poisson bracket between the primary and the secondary constraint vanishes:
\begin{align}
	\{\Pi^{\red{0}}, \partial_a \Pi^{a}\} = 0\,.
\end{align}
This prevents the spatial derivative from being moved onto a Lagrange multiplier. In GR, one encounters a very similar but slightly more complicated situation. As is well-known, the momentum densities conjugate to lapse and shift both vanish,
\begin{align}
	\tilde{\pi}_0 &\approx 0 &\text{and} && \tilde{\pi}_a &\approx 0\,.
\end{align}
These constraints give rise to secondary constraints, namely the so-called scalar and vector constraints,
\begin{align}
	\tilde{\mathcal{S}} &= \frac{1}{\sqrt{h}}\left(\tilde{\pi}_{ab}\tilde{\pi}^{ab} - \frac12 \tilde{\pi}\ud{a}{a} \,\tilde{\pi}\ud{b}{b}\right) - \sqrt{h}\, ^{(3)}\mathcal{R} \approx 0\notag\\
	\tilde{\mathcal{V}}_a &= 2\mathfrak{D}_b\tilde{\pi}\ud{b}{a} \approx 0\,.
\end{align}
Both of these constraints contain spatial derivatives. The spatial derivatives of the scalar constraints are hidden inside $^{(3)}\mathcal{R}$, which contains terms of the form $\partial_c h_{ab}$, while the spatial derivatives of the vector constraint are plainly visible. Just as in electromagnetism, this does not cause any issues. However, the reason is slightly different. Whereas in electromagnetism the Poisson bracket between the primary and the secondary constraint is exactly zero, in GR one finds for the smeared primary and secondary constraints
\begin{align}
	\{\pi_\mu[f_1], \pi_\nu[f_2]\} &= 0 & \text{for } \mu,\nu \in\{0, a\}\notag\\
	\{\pi_\mu[f_1], \mathcal{S}[f_2]\} &= 0 & \text{for } \mu\in\{0, a\}\notag\\
	\{\pi_\mu[f] , \mathcal{V}[\vec{v}]\} &= 0 & \text{for } \mu\in\{0, a\}\notag\\
	\{\mathcal{S}[f], \mathcal{V}[\vec{v}]\} &= - \mathcal{S}[\mathcal{L}_{\vec{v}}f] \approx 0 \notag\\
	\{\mathcal{V}[\vec{v}_1], \mathcal{V}[\vec{v}_2]\} &= \mathcal{V}[\mathcal{L}_{\vec{v}_1}\vec{v}_2] \approx 0 \notag\\
	\{\mathcal{S}[f_1], \mathcal{S}[f_2]\} &= \mathcal{V}[\left(f_1\partial^{a}f_2 - f_2 \partial^{a}Êf_1\right) \partial_a] \approx 0\,,
\end{align}
where $f_i$ and $\vec{v}_i$ are test functions. Notice that the Poisson brackets between the primary and secondary constraints vanish identically, while the Poisson brackets between the secondary constraints vanish on the constraints surface for any choice of test functions. Since in Step~\red{5} of the Dirac-Bergmann algorithm the consistency condition is evaluated on the constraint surface, this is sufficient to guarantee that the field theoretic analogue of equation~\eqref{eq:ConsistencyCondition} does not lead to differential equations for the Lagrange multipliers. 

These two examples clearly teach us that, as already stated, the presence of spatial derivatives in constraints is not sufficient to conclude that the Dirac-Bergmann algorithm will fail. If the Poisson brackets between the constraints vanish exactly or on the constraint surface, then the algorithm stands a chance of succeeding (it is still possible that the consistency condition generates a new constraint with spatial derivatives and which has a non-vanishing Poisson bracket with at least one of the old constraints, thus leading to a failure at a later point in the algorithm).


\section{An Upper and Lower Bound on the Degrees of Freedom of $f(\Q)$ Gravity}\label{sec:UpperBound}
\subsection{Deriving a Lower Bound from Cosmology and an Upper Bound from the Kinetic Matrix}\label{ssec:KineticMatrixApproach}
Even though the Hamiltonian analysis of $f(\Q)$ gravity fails to answer the question of how many physical degrees of freedom propagate, it is still possible to give an upper and lower bound for this number.

A lower bound is actually already known since the first work on $f(\Q)$ cosmology~\cite{BeltranJimenez:2019tme}. Using a perturbative argument, it was shown that $f(\Q)$ gravity propagates \textit{at least} two additional degrees of freedom compared to standard GR. Thus, the lower bound for the total number of physical degrees of freedom is four.

To obtain an upper bound for this number, we will analyze the so-called \textit{kinetic matrix} of the theory. We recall that given a system of second order partial differential equations for $N$ fields $\Psi \ce (\Psi_1(t,x), \dots, \Psi_N(t, x))^T$, the kinetic matrix $\mathcal{K}$ is the matrix which multiplies the second order time derivatives, $\ddot{\Psi}$. By determining the rank $R\ce \text{rank}(\mathcal{K})$  we can infer the smallest amount of constraints that are present in the theory, which in turn allows us to give an upper bound on the number of constraints and degrees of freedom, namely $N-R$ and $R$, respectively. These facts are explained in more detail in Appendix~\ref{app:PDEs}. 

Here, we dive right into the determination of $\mathcal{K}$ and its rank in the case of $f(\Q)$ gravity. To begin with, we recall that the field equations of $f(\Q)$ gravity can be written as~\cite{DAmbrosio:2021, DAmbrosio:2021b}
\begin{align}
		f'(\Q)\, G_{\mu\nu} - \frac12 \left(f(\Q) - f'(\Q)\, \Q\right)g_{\mu\nu} + 2f''(\Q)\, P\ud{\alpha}{\mu\nu}\partial_\alpha\Q &= T_{\mu\nu} \label{eq:f(Q)MetricFieldEquations} \\
		\nabla_\mu \nabla_\nu \left(\sqrt{|g|}\, f'(\Q) \, P\ud{\mu\nu}{\alpha}\right) &= 0 \label{eq:f(Q)ConnectionFieldEquations}\,.
\end{align}
The first set of equations is referred to as \textit{metric field equations}, while the second set are the \textit{connection field equations}. The tensor $G_{\mu\nu}$ is the standard Einstein tensor with respect to the Levi-Civita connection, $G_{\mu\nu} \ce \R_{\mu\nu} - \frac12 \R\, g_{\mu\nu}$, while the non-metricity conjugate reads\footnote{This follows from~\eqref{eq:NonMetricityConjugate} for the parameter choice~\eqref{eq:STEGRParameters}.}
\begin{align}
	P\ud{\alpha}{\mu\nu} = -\frac14 Q\ud{\alpha}{\mu\nu} + \frac12 Q\dud{(\mu}{\alpha}{\nu)} + \frac14 g_{\mu\nu}Q^\alpha -\frac14 \left(g_{\mu\nu}\bar{Q}^\alpha + \delta\ud{\alpha}{(\mu}Q_{\nu)}\right),
\end{align}
and $\Q$ is the non-metricity scalar as defined in~\eqref{eq:NonMetricityScalar}. Recall that the affine connection~$\G{\alpha}{\mu\nu}$ is postulated to be flat and torsionless, which means that it can be written as
\begin{align}
	\G{\alpha}{\mu\nu} = \PD{x^\alpha}{\xi^\lambda} \partial_\mu \partial_\nu \xi^\lambda\,,
\end{align}
where $\xi^\alpha$ are four arbitrary functions of the spacetime coordinates. Notice that the metric field equations are symmetric in $\mu$ and $\nu$ and that the connection field equations have only one free index, namely $\alpha$. Thus, we have a total of $10+4$ field equations for $10$ metric components and $4$ connection functions $\xi^\alpha$.

Since these are tensorial equations, we have the freedom to apply diffeomorphisms which act simultaneously on $\xi^\alpha$ and $g_{\mu\nu}$. This realisation helps us in simplifying the discussion, because there is a particularly simple choice of gauge: The coincident gauge. As we have explained in section~\ref{sec:TeleparallelTheories}, the coincident gauge is always a viable choice within any theory based on a flat and torsionless connection, independent of any action principle or field equations. In using this gauge freedom, we have essentially fixed $\xi^\alpha$ and we are left with $10+4$ equations for the ten metric components.

It seems that now we run into the problem of having an over-determined system. However, this is not the case. Because the action functional~\eqref{eq:f(Q)Action} which defines $f(\Q)$ gravity is diffeomorphism invariant, one can show that the following Bianchi identities hold~\cite{BeltranJimenez:2020}:
\begin{align}\label{eq:BianchiIdentity}
	\D_\alpha \M\ud{\alpha}{\mu} + \mathcal{C}_\mu = 0\,,
\end{align}
where $\mathcal{D}$ is the covariant derivative operator with respect to the Levi-Civita connection,~$\mathcal{M}_{\mu\nu}$ stands for the metric field equations (i.e., the left hand side of~\eqref{eq:f(Q)MetricFieldEquations}), and~$\mathcal{C}_\mu$ stands for the connection field equations (i.e., the left hand side of~\eqref{eq:f(Q)ConnectionFieldEquations}).

We conclude that when the ten metric field equations are satisfied, $\M_{\mu\nu} = 0$, the connection field equations are automatically satisfied by virtue of the Bianchi identity~\eqref{eq:BianchiIdentity}. It follows that the use of the coincident gauge ``moves'' all physical degrees of freedom into the ten metric components and that we only need to study the metric field equations in order to determine these degrees of freedom.

It is convenient to introduce a foliation of $(\M, g_{\mu\nu})$, as explained in subsection~\ref{ssec:DiracBergmannAnalysis}, and to work with the ADM variables $\{N, N^{a}, h_{ab}\}$. Our task is now to isolate all second order time derivatives of lapse $N$, shift $N^{a}$, and the intrinsic metric $h_{ab}$ in the metric field equations~\eqref{eq:f(Q)MetricFieldEquations}. Given the above form of the field equations, this is straightforward:
\begin{itemize}
	\item[(a)] The non-metricity scalar $\Q$ is constructed solely from the non-metricity tensor $Q_{\alpha\mu\nu}\ce \nabla_\alpha g_{\mu\nu}$ and contains therefore only first order derivatives of the ADM variables.
	\item[(b)] The first term of the metric field equations, $f'(\Q)\, G_{\mu\nu}$, contains only second order time derivatives of $h_{ab}$, but not of $N$ and $N^{a}$. This is a well-known property of the Einstein tensor.
	\item[(c)] The second term, $-\frac12\left(f(\Q) - f'(\Q)\, \Q\right)g_{\mu\nu}$, is highly non-linear but it contains only first order derivatives of all variables.
	\item[(d)] The third term, $2 f''(\Q) P\ud{\red{0}}{\mu\nu}\partial_{\red{0}} \Q$, contains second order time derivatives of all ADM variables. (Here we have omitted the terms $P\ud{a}{\mu\nu}\partial_a \Q$ since they are irrelevant to the present discussion).
\end{itemize}
Thus, the kinetic matrix $\mathcal{K}$ receives contributions from $G_{\mu\nu}$ and from $P\ud{\red{0}}{\mu\nu}\partial_{\red{0}} \Q$. Given this fact, it is convenient to write the kinetic matrix as the sum of two matrices, 
\begin{align}
	\mathcal{K}\ce f'(\Q)\,\mathcal{K}_\text{GR} + 2 f''(\Q)\,\mathcal{K}_\text{M},
\end{align}
where $\mathcal{K}_\text{GR}$ is the kinetic matrix of GR and $\mathcal{K}_\text{M}$ encodes the modifications to the dynamics coming from $P\ud{\red{0}}{\mu\nu}\partial_{\red{0}} \Q$.

To explicitly construct these matrices, we choose the following conventions: The second order time derivatives of our basic metric variables are collected into a column vector $\ddot{\Psi}$, defined as
\begin{align}
	\ddot{\Psi} \ce (\ddot{N}, \ddot{N}^{1}, \ddot{N}^{2}, \ddot{N}^{3}, \ddot{h}_{11}, \ddot{h}_{12}, \ddot{h}_{13}, \ddot{h}_{22}, \ddot{h}_{23}, \ddot{h}_{33})^\transpose\,.
\end{align}
The $10\times 10$ matrix $\mathcal{K}$ acts on $\ddot{\Psi}$ and generates the terms containing second order time derivatives in the metric field equations. The components of the $10$-dimensional vector $\mathcal{K}\, \ddot{\Psi}$ correspond to the $00$, $01$, $02$, $03$, $11$, $12$, $13$, $22$, $23$, $33$ components (in this order) of the field equations. It is now evident that $\mathcal{K}_\text{GR}$ has the form
\begin{align}
	\mathcal{K}_\text{GR} = 
	\left(\begin{array}{c|c}
		0_{4\times 4} & 0_{4\times 6} \\ \hline
		0_{6\times 4} & D_{6\times 6}
	\end{array}\right)\,,
\end{align}
where $0_{m\times n}$ denotes a matrix of dimension $m\times n$ whose entries are all zero, and $D_{6\times 6}$ is a placeholder for the actual non-trivial $6\times 6$ matrix which describes the second order time derivatives of $h_{ab}$ in the field equations. The upper left and upper right blocks of $\mathcal{K}_\text{GR}$ simply express the fact that there are no second order times derivatives in the components $G_{\red{0}\mu}$, while the lower left block tells us that there are no second order time derivatives of lapse and shift in the space-space components $G_{ab}$. A direct inspection of the space-space components of the Einstein tensor finally reveals that the lower right block matrix in $\mathcal{K}_\text{GR}$ is generated by the term
\begin{align}
	\frac{1}{2N}\left(\delta\du{m}{a}\delta\du{n}{b} - h^{ab}h_{mn}\right)\,,
\end{align}
which, using our conventions, translates into the matrix
\begin{align}
	D = \frac{1}{2N}
	\begin{pmatrix}
		1-h^{11}h_{11} & h^{12}h_{11} & h^{13}h_{11} & h^{22}h_{11} & h^{23}h_{11} & h^{33}h_{11} \\
		h^{11}h_{12} & 1-h^{12}h_{12} & h^{13}h_{12} & h^{22}h_{12} & h^{23}h_{12} & h^{33}h_{12} \\
		h^{11}h_{13} & h^{12}h_{13} & 1 - h^{13}h_{13} & h^{22}h_{13} & h^{23}h_{13} & h^{33}h_{13} \\
		h^{11}h_{22} & h^{12}h_{22} & h^{13}h_{22} & 1 - h^{22}h_{22} & h^{23}h_{22} & h^{33}h_{22} \\
		h^{11}h_{23} & h^{12}h_{23} & h^{13}h_{23} & h^{22}h_{23} & 1 - h^{23}h_{23} & h^{33}h_{23} \\
		h^{11}h_{33} & h^{12}h_{33} & h^{13}h_{33} & h^{22}h_{33} & h^{23}h_{33} & 1 - h^{33}h_{33}
	\end{pmatrix}\,,
\end{align}
which has rank six. In order to determine the matrix $\mathcal{K}_\text{M}$, it is convenient to use the result obtained in~\cite{DAmbrosio:2020}, namely that the non-metricity scalar $\Q$ in terms of ADM variables can be written as
\begin{align}\label{eq:QinADMVariables}
	\Q = \pi\, \dot{N} + \pi_a\, \dot{N}^{a} + \pi^{ab}\dot{h}_{ab} + \text{terms without time derivatives}\,,
\end{align}
where the conjugate momenta are explicitly given by\footnote{The momenta given here differ from the momenta given in~\eqref{eq:MomentumDensities} even for the case $\phi=1$ because here we did not use any integrations by parts to simplify $\Q$. In fact, we are not allowed to, since we are merely rewriting $\Q$ in terms of ADM variables, rather than the action itself.}
\begin{align}
	\pi &\ce \PD{\Q}{\dot{N}} = -\frac{1}{N^3}\partial_a N^{a} \notag\\
	\pi_a &\ce \PD{\Q}{\dot{N}^{a}} = -\frac{1}{N^3}\left(\frac12N\,h^{bc}\partial_a h_{bc} - \partial_a N\right) \notag\\
	\pi^{ab} &\ce \PD{\Q}{\dot{h}_{ab}} = \frac{1}{N}\left(K^{ab} - h^{ab} K\ud{c}{c}\right) +\frac{1}{2N^2} h^{ab}\partial_c N^{c}\notag\\
	&\phantom{:}= \frac{1}{2N^2}\left(h^{ac}h^{bd} - h^{ab}h^{cd}\right)\dot{h}_{cd} + \frac{1}{2N^2}\left(h^{ab}h^{de} - h^{ad}h^{be}\right)N^{c}\partial_c h_{de}\notag\\
	&\phantom{\ce}\ + \frac{1}{2N^2}\left(h^{ab} \partial_c N^{c} - h^{bc}\partial_c N^{a} - h^{ac}\partial_c N^{b}\right)\,.
\end{align}
Observe that the momenta conjugate to lapse and shift do \textit{not} depend on any velocities, while $\pi^{ab}$ is linear in $\dot{h}_{cd}$. Therefore, the action of the differential operator $P\ud{\red{0}}{\mu\nu}\partial_{\red{0}}$ on $\Q$ reduces to
\begin{align}
	P\ud{\red{0}}{\mu\nu}\partial_{\red{0}}\Q &=  P\ud{\red{0}}{\mu\nu}\left(\pi\, \ddot{N} + \pi_a\, \ddot{N}^{a} + \rho^{ab}\, \ddot{h}_{ab}\right) \notag\\
	&\phantom{=}\ + \text{terms involving first order time derivatives}\,,
\end{align}
where we have introduced
\begin{align}
	\rho^{ab} \ce \pi^{ab} + \underbrace{\frac{1}{2N^2}\left(h^{ac} h^{bd} - h^{ab} h^{cd}\right)\dot{h}_{cd}}_{\subset \dot{\pi}^{ab}}\,.
\end{align}
In this form, it is particularly simple to read off the matrix $\mathcal{K}_\text{M}$. First off, one notices that the ten rows of~$\mathcal{K}_\text{M}$ can be written as a single row vector multiplied by ten different pre-factors. Concretely, one finds
\begin{align}
	\mathcal{K}_\text{M} =
	\begin{pmatrix}
		P\ud{\red{0}}{\red{00}} R \\
		P\ud{\red{0}}{\red{0}1} R \\
		P\ud{\red{0}}{\red{0}2} R \\
		P\ud{\red{0}}{\red{0}3} R \\
		P\ud{\red{0}}{11} R \\
		P\ud{\red{0}}{12} R \\
		P\ud{\red{0}}{13} R \\
		P\ud{\red{0}}{22} R \\
		P\ud{\red{0}}{23} R \\
		P\ud{\red{0}}{33} R
	\end{pmatrix}\,,
\end{align}
where the row vector $R$ is explicitly given by
\begin{align}
	R \ce \begin{pmatrix}
		\pi & \pi_1 & \pi_2 & \pi_3 & \rho^{11} & \rho^{12} & \rho^{13} & \rho^{22} & \rho^{23} & \rho^{33}
	\end{pmatrix}\,.
\end{align}
This immediately tells us that $\mathcal{K}_\text{M}$ has less than full rank. In fact, since all rows are linearly dependent, the rank of $\mathcal{K}_\text{M}$ is just one!

We can now put together all pieces and we find that the kinetic matrix can schematically be written as
\begin{align}
	\mathcal{K} = f'(\Q)\, \mathcal{K}_\text{GR} + 2 f''(\Q)\,\mathcal{K}_\text{M} \quad\longrightarrow\quad
	\left(\begin{array}{c|c}
		2 f''(\Q) R_\pi  & 2 f''(\Q) R_\rho \\
		0_{3\times 4} & 0_{3 \times 6} \\ \hline
		0_{6\times 4} & f'(\Q)\,D_{6\times 6}
	\end{array}\right)\,,
\end{align}
where $R_\pi$ and $R_\rho$ denote the row vector composed of the first four and last six entries of $R$, respectively, and the arrow symbolizes the process of eliminating linearly dependent rows, which is a necessary step in the determination of $\mathcal{K}$'s rank. If $f''(\Q)=0$ with $f'(\Q)\neq 0$, the contribution of $\mathcal{K}_\text{M}$ is absent and the kinetic matrix has rank six. This is what one would expect, since the case  $f''(\Q)=0$ with $f'(\Q)\neq 0$ simply corresponds to recovering GR and rank six means there are $10-6=4$ constraints. These are precisely the constraints that arise from having non-dynamical lapse and shift fields.

The more interesting case is when $f''(\Q)\neq 0$, since this leads to modifications. In this case it follows that the rank of $\mathcal{K}$ is seven, which conversely implies the presence of three constraints and it places an upper bound on the degrees of freedom: There are at most seven propagating degrees of freedom in $f(\Q)$ gravity. 

In conclusion, we point out that the analysis performed in this section is completely independent of the Hamiltonian analysis which we attempted in section~\ref{sec:DiracBergmannAlgorithm}. The presence of partial differential equations for the Lagrange multipliers we encountered there cast doubt on the results obtained in~\cite{Hu:2022}. The analysis of this section finally conclusively shows that the result obtained in~\cite{Hu:2022}, namely that there are eight degrees of freedom, can not be correct.

\subsection{On a Subtlety Related to Integration by Parts in $f(\Q)$ Gravity}\label{sec:Subtlety}
It might seem puzzling at first that the non-metricity scalar~\eqref{eq:QinADMVariables} written in ADM variables is only linear in $\dot{N}$ and $\dot{N}^{a}$, while the metric field equations~\eqref{eq:f(Q)MetricFieldEquations}, as we have shown in the previous subsection, contain the second order time derivatives $\ddot{N}$ and $\ddot{N}^{a}$. One would naively expect that a Lagrangian which is quadratic in some velocities but only linear in others does not lead to field equations which contain second order time derivatives of all variables. In fact, in STEGR, where the Lagrangian is $\mathcal{L} = \sqrt{-g}\, \Q$, the field equations contain second order time derivatives of $h_{ab}$, but \textit{not} of $N$ and $N^{a}$. This is consistent with the usual expectations.

Why then do we obtain second order time derivatives of all variables in $f(\Q)$ gravity? The simple and qualitative answer is that even tough $\dot{N}$ and $\dot{N}^{a}$ appear only linearly in $\Q$, the Lagrangian $\mathcal{L} = \sqrt{-g}\, f(\Q)$ is in general a \textit{non-linear function} of $\Q$. Thus, in general,~$\dot{N}$ and $\dot{N}^{a}$ do in fact \textit{not} appear linearly in the Lagrangian and it has to be expected that the field equations contain second order time derivatives of lapse and shift.

Quantitatively, this can be seen as follows: The Lagrangian of $f(\Q)$ gravity can schematically be written as 
\begin{align}
	\mathcal{L} = N\, \sqrt{h}\, f(\Q) \equiv N\, \sqrt{h}\, f(N, \dot{N}, \dots)\,,
\end{align}
where the dots stand collectively for the other ADM variables and their derivatives. If we vary the action
\begin{align}
	\S[N, \dots] \propto \int_{\mathcal{M}}\dd^4 x\, N\, \sqrt{h}\, f(N, \dot{N}, \dots)
\end{align}
with respect to $N$, we obtain
\begin{align}
	\delta_N \S[N, \dots] \propto \int_\M \dd^4 x\, \sqrt{h} \left(f\,\delta N + N\, f'\,\delta N + N\,f'\, \delta\dot{N} + \dots \right)\,,
\end{align}
where $f'$ stands for $\frac{\dd f}{\dd \Q}$, as always. In order to properly derive the field equations, we have to perform an integration by parts which removes the time derivative from $\delta\dot{N}$. As usual, we can drop boundary terms arising from this integration since~$\delta N$ vanishes on the boundary and we simply find
\begin{align}
	\delta_N \S[N, \dots] &\propto \int_\M \dd^4 x\, \sqrt{h} \left(f\,\delta N + N\, f'\,\delta N - \frac{\dd}{\dd t}\left[ N\,f'\right]\, \delta N + \dots\right)\,\notag\\
	&= \int_\M \dd^4 x\, \sqrt{h}\left(f + N\, f' - \dot{N}\, f' - N\, f''\, \dot{\Q} + \dots\right)\delta N\,.
\end{align}
By demanding that this variation vanishes for all $\delta N$ in the bulk, we obtain the field equation
\begin{align}
	f + N\, f' - \dot{N}\, f' - N\, f''\, \dot{\Q} + \dots \phantom{.}=\phantom{.} 0\,,
\end{align}
which contains second order time derivatives of $N$ because of the term $f''\, \dot{\Q}$. It is also evident that this is due to the non-linear nature of $f$, as explained before. For $f''=0$, i.e., if $f$ is a linear function, then the second order time derivatives disappear from the field equations.


\section{Conclusion}\label{sec:Conclusions}
The concept of symmetry lies at the very core of theoretical physics, although the specific geometric representation employed to illustrate it is subject to conventional choices. The true triumph of GR never resided solely in the geometrization of gravity, but rather in its unification with inertia. This unification, epitomized by the equivalence principle, signifies that gravity can always be locally eradicated through a change in coordinate systems. Simultaneously, a fundamental characteristic of gauge theories emerges, wherein a gauge field force can be made to locally vanish when its field strength is zero. We may ponder that the coincident GR, which purges both torsion and curvature from gravity, could be the theory of gravity, embracing the astounding accomplishments of the gauge principle in the realm of particle physics.
This formulation of GR based on non-metricity represents also a unique extension of gravity with interesting cosmological and astrophysical implications. Theories based on $f(\Q)$ give rise to accelerating universes and hairy black hole solutions. As a modification of GR, it naturally contains additional degrees of freedom. How many these are still remains an open question.  Cosmological perturbation analysis puts a lower bound of four degrees of freedom. Previous studies in the literature unfortunately blindly applied the Dirac-Bergmann algorithm, not realising that $f(\Q)$ gravity breaks one of the basic assumptions of the algorithm, and erroneously claimed that the theory propagates eight degrees of freedom. In this work, we explicitly showed that the Hamiltonian analysis based on the Dirac-Bergmann algorithm fails in $f(\Q)$ gravity. We have identified the reason for this failure and we argued that other teleparallel theories of gravity based on either torsion or non-metricity likely suffer from the same shortcoming. As an alternative approach, we studied the kinetic matrix of~$f(\Q)$ gravity in ADM variables. This allowed us to put an upper bound of seven degrees of freedom.


\section*{Acknowledgments}
LH is supported by funding from the European Research Council (ERC) under the European Unions Horizon 2020 research and innovation programme grant agreement No 801781 and by the Swiss National Science Foundation grant 179740. LH further acknowledges support from the
Deutsche Forschungsgemeinschaft (DFG, German Research Foundation) under Germany's Excellence Strategy EXC 2181/1 - 390900948 (the Heidelberg STRUCTURES Excellence Cluster).

This research was supported in part by the Perimeter Institute for Theoretical Physics. Research at Perimeter Institute is supported by the Government of Canada through the Department of Innovation, Science and Economic Development and by the Province of Ontario through the Ministry of Colleges and Universities.
\newpage

\appendix
\section{Elements of the Mathematical Theory of 1$^\text{st}$ and 2$^\text{nd}$ Order PDEs}\label{app:PDEs}
In the main text we made use of two tools stemming from the theory of partial differential equations (PDEs): The condition~\eqref{eq:SolvabilityCriterion}, which tells us when a first order PDE has a unique solution, and the kinetic matrix and its various properties. In this appendix we provide the necessary mathematical background to comprehend these tools and how to work with them, however, without being mathematically too precise. Intuition plays a more central role for us here. More details on the subject of PDEs and, in particular, a rigorous formulation which can withstand the scrutiny of the mathematically inclined reader can be found for instance in the book~ \cite{MiersemannBook}.

\subsection{First Order PDEs and the Solvability Criterion}
We are concerned with first order PDEs for $N$ variables in $D$ spacetime dimensions. Let us denote the $N$ variables by the $N$-dimensional vector $\Psi \ce (\Psi_1, \dots, \Psi_N)^\transpose$. Its components are functions of the $D$ spacetime coordinates $x^\mu = (x^{1}, \dots, x^{D})$. What we are interested in is the so-called \textit{initial value problem for first order PDEs}. Specifically, we want to prescribe initial data on a $(D-1)$-dimensional \textit{Cauchy} or \textit{initial value} hypersurface $\S$ and answer the question, whether a given PDE together with initial data on $\S$ uniquely determines $\Psi$ everywhere in the spacetime. Qualitatively speaking, if we have a first order PDE, we have to integrate once to find $\Psi$, which means there will be one ``integration constant''. Thus, the initial data we have to prescribe in order to fix the ``integration constant'' is the ``value'' of~$\Psi$ on $\S$. We write this symbolically as $\left.\Psi\right|_{\S} = F$, where $F$ is a vector field which represents the data we have chosen. 

From a physicist's perspective, we can think of the initial value problem as follows: We are interested in a certain field $\Psi$ and we assume we have measured, or that we know through some other means, that at a certain instant of time this field is given by $F$. This instant of time is the $(D-1)$-dimensional hypersurface $\S$. Given this knowledge, can we predict how $\Psi$ will evolve into the future of $\S$? Similarly, can we retrodict how $\Psi$ behaved in the past of $\S$? In other words, can we determine $\Psi$ everywhere in the spacetime given that we know its field equations and given that we know that on $\S$ it is equal to $F$?

There is a surprisingly simple criterion, which we will derive. The starting point is the fact that so-called \textit{quasi-linear first order partial differential equations}\footnote{A quasi-linear PDE is a PDE whose highest order derivative appears \textit{linearly}. Note, however, that if the highest order derivative is of order $n$, this derivative can be multiplied by a function of $\Psi$ and its first $n-1$ derivatives, without changing the fact that the $n$-th order derivative appears linearly. See~\cite{MiersemannBook} for details.} for $\Psi$ can be compactly written as
\begin{align}\label{eq:InitialValueProblem}
	\begin{cases}
		\displaystyle\sum_{i=1}^{D} M^{(i)}(x)\,\partial_i\Psi + L(x)\, \Psi + V(x) = 0 &\\
		\left.\Psi\right|_{\S} = F
	\end{cases}\,,
\end{align}
where $M^{(i)}$ with $i\in\{1,\dots, D\}$ are $D$ matrices of dimension $N\times N$, $L$ is also a $N\times N$ matrix, and $V$ is a $N$-dimensional vector. All these matrices and the vector $V$ are allowed to depend on the field $\Psi$, but not on its derivatives. Since the components of $\Psi$ are functions of the spacetime coordinates $x^\mu$, we simply abbreviate the dependence of the above matrices and vectors by $x$. To proceed, we need two key ideas:
\begin{enumerate}
	\item It is always possible to define the Cauchy surface $\S$ implicitly by $\rchi(x) = 0$, where $\rchi$ is some scalar function which satisfies $\nabla\rchi(x) \neq 0$ for all $x$. Here, $\nabla\rchi$ simply denotes the gradient of $\rchi$. In physics one often defines $\S$ as the $t=const.$ surface, which means it is given by $\rchi(x) = t- const. = 0$, and the condition $\nabla\rchi(x) \neq 0$ simply means that $\S$ has a nowhere vanishing normal vector $n\ce \nabla\rchi$. 
	\item The initial value problem~\eqref{eq:InitialValueProblem} can be written in coordinates which are ``adapted'' to the problem. That is, we can introduce a coordinate system such that $\partial_{1}\Psi$, $\partial_2\Psi$, and so on are all known, except $\partial_D\Psi$. The PDE~\eqref{eq:InitialValueProblem} can then be solved for $\partial_D\Psi$ and the solution $\Psi$ is found by integrating $\partial_D \Psi$.
\end{enumerate} 
In order to introduce these adapted coordinates, we define a diffeomorphism $\phi:\M\to\M$ and demand that its $D$-th component is given by
\begin{align}
	\phi^D(x) \ce \rchi(x)\,.
\end{align}
Furthermore, since $\phi$ is a diffeomorphism, it holds true that the Jacobian matrix $J$ of the change of coordinates, i.e., $J\ud{\mu}{\nu}\ce \PD{\phi^\mu}{x^\nu}$ with $\mu,\nu\in\{1,\dots, D\}$, has a non-zero determinant. 

Let us briefly pause and put the situation thus far into simple words: We can represent the Cauchy surface $\S$ by the coordinate constraint equation $\rchi(x) = 0$. The condition $\nabla\rchi(x) \neq 0$ ensures that $\S$ has a non-vanishing normal vector $n\ce \nabla\rchi$. In the new coordinates $\phi^\mu = (\phi^1, \dots, \phi^D)$, the surface $\S$ is simply described by $\phi^D = 0$. It thus follows that the coordinates $(\phi^1,\dots, \phi^{D-1})$ all lie within $\S$. A change in the coordinate $\phi^D$ results in a ``movement'' in the direction orthogonal to $\S$, i.e., a movement along the normal vector~$n$. 

We now proceed in re-writing the PDE~\eqref{eq:InitialValueProblem} in these new coordinates. To that end, we need
\begin{align}
	\PD{\Psi}{x^{\mu}} = \PD{\Theta}{\phi^\nu}\PD{\phi^\nu}{x^\mu}\,,
\end{align}
where we introduced $\Theta(\phi) \ce \Psi(\phi(x))$. This simple relation immediately tells us that from our knowledge of the initial data $F$ on $\S$ we can determine the partial derivatives $\PD{\Theta}{\phi^{a}}$ with $a\in\{1,\dots, D-1\}$, when evaluated on $\S$. The only partial derivative which remains undetermined by the initial data is $\PD{\Theta}{\phi^{D}}$. This is seen by recalling the definition of partial derivatives evaluated on $\S$ (i.e., evaluated on $\phi^{D} = 0$):
\begin{align}
	\left.\PD{\Theta}{\phi^{a}} \right|_{\S} &\ce \lim_{\epsilon\to 0}\frac{\Theta(\phi^1,\dots, \phi^{a} + \epsilon,\dots, \phi^{D-1},0) - \Theta(\phi^1,\dots, \phi^{a},\dots, \phi^{D-1},0)}{\epsilon} \notag\\
	&= \lim_{\epsilon\to 0} \frac{F(\phi^1,\dots, \phi^{a} + \epsilon,\dots, \phi^{D-1},0) - F(\phi^1,\dots, \phi^{a},\dots, \phi^{D-1},0)}{\epsilon} \notag\\
	&= \PD{F}{\phi^{a}}\,,
\end{align}
which holds for all $a\in\{1,\dots, D-1\}$. Thus, these partial derivatives are all determined by the data~$F$, as claimed. However, the partial derivative $\PD{\Theta}{\phi^D} = \PD{\Theta}{\rchi}$ remains undetermined because
\begin{align}\label{eq:DThetaDPhiD}
	\left.\PD{\Theta}{\rchi}\right|_{\S} &= \lim_{\epsilon\to 0}\frac{\Theta(\phi^1,\dots, \phi^{a},\dots, \phi^{D-1},\epsilon) - \Theta(\phi^1,\dots, \phi^{a},\dots, \phi^{D-1},0)}{\epsilon} \notag\\
	&=\lim_{\epsilon\to 0}\frac{\Theta(\phi^1,\dots, \phi^{a},\dots, \phi^{D-1},\epsilon) - F(\phi^1,\dots, \phi^{a},\dots, \phi^{D-1},0)}{\epsilon}\,,
\end{align}
i.e., because we do not know the value of $\Theta(\phi^1,\dots, \phi^{a},\dots, \phi^{D-1},\epsilon)$, since $\Theta$ is evaluated at a point which is \textit{not} on $\S$. 

Equation~\eqref{eq:DThetaDPhiD} also gives us another insight: If we are able to determine $\left.\PD{\Theta}{\rchi}\right|_{\S}$, then we can determine $\Theta$ in a neighbourhood of $\S$. In fact, if we know $\left.\PD{\Theta}{\rchi}\right|_{\S}$, then it follows from~\eqref{eq:DThetaDPhiD} that
\begin{align}
	\Theta(\phi^1,\dots, \phi^{a},\dots, \phi^{D-1},\epsilon) = F + \epsilon\, \left.\PD{\Theta}{\rchi}\right|_{\S} + \mathcal{O}(\epsilon^2)\,.
\end{align}
Thus, we can formally integrate the equation and determine $\Theta$ also away from $\S$. This is where the PDE~\eqref{eq:InitialValueProblem} enters the game. In fact, in the new coordinate system, we can re-write the PDE as
\begin{align}
	\sum_{i=1}^{D} \tilde{M}^{(i)} \PD{\rchi}{x^i}\PD{\Theta}{\rchi} + \sum_{i=1}^{D} \tilde{M}^{(i)} \PD{\phi^{a}}{x^{i}}\PD{\Theta}{\phi^{a}} + \tilde{L}\, \Theta + \tilde{V} = 0\,,
\end{align}
where $\tilde{M}^{(i)}$, $\tilde{L}$, and $\tilde{V}$ are the same matrices and vectors as before, but expressed in the new coordinate system. Now observe that if we evaluate this equation on $\S$, we obtain schematically
\begin{align}
	\left.\sum_{i=1}^{D} \tilde{M}^{(i)} \PD{\rchi}{x^i}\PD{\Theta}{\rchi}\right|_{\S} = \text{terms known on }\S\,.
\end{align}
It follows that if we are able to solve this equation for $\left.\PD{\Theta}{\phi^{D}}\right|_{\S}$, we can integrate the PDE~\eqref{eq:InitialValueProblem} and determine $\Theta$ (which is the same as determining $\Psi$) also away from the surface $\S$. Thus, determining whether the initial value problem~\eqref{eq:InitialValueProblem} has a unique solution or not is reduced to a problem of linear algebra: One has to determine whether the matrix $\sum_{i=1}^{D} \tilde{M}^{(i)} \PD{\rchi}{x^i}$ is invertible. 

Recall that $\S$ has a normal vector $n\ce \nabla\rchi$, whose components are simply given by $n_i =\PD{\rchi}{x^{i}}$ for $i\in\{1,\dots, D\}$. Furthermore, a matrix is invertible if and only if its determinant is not zero. It follows that the initial value problem admits a unique solution in a neighbourhood of $\S$ if an only if
\begin{align}\label{eq:1stOrderSolvabilityCriterion}
	\det\left(\sum_{i=1}^{D} \tilde{M}^{(i)}(x) \, n_i(x)\right) &\neq 0  &\text{for all } x\in\S\,.
\end{align}
This is precisely the solvability criterion~\eqref{eq:SolvabilityCriterion} we have used when analyzing the PDEs~\eqref{eq:SchematicFormPDEs} for the Lagrange multipliers, which emerged from the Dirac-Bergmann algorithm. In~\eqref{eq:SolvabilityCriterion}, we found that for that particular PDE the determinant is zero for all $x$, no matter how one chooses the Cauchy surface. Hence, it is never possible to find a unique solution to that PDE. One can show that the rank of the matrix $\sum_{i=1}^{3}M^{(i)}n_i$ is two, which means that after performing linear operations on the PDE~\eqref{eq:SchematicFormPDEs} (such as adding/subtracting equations from each other or multiply/divide them by certain functions) one can bring the PDE into the form
\begin{align}
	\begin{pmatrix}
		1 & * & * \\
		0 & 1 & * \\
		0 & 0 & 0
	\end{pmatrix}
	\begin{pmatrix}
		\partial_z \lambda^{1} \\
		\partial_z \lambda^{2} \\
		\partial_z \lambda^{3}
	\end{pmatrix}
	 + \text{ terms known on }\S = 0\,.
\end{align}
Evidently, in order to determine $\partial_z \lambda^{1}$ and $\partial_z \lambda^{2}$, one has to prescribe $\partial_z \lambda^{3}$ by hand. This holds in full generality: If the rank of the matrix $\sum_{i=1}^{D}\tilde{M}^{(i)}n_i$ is $r<D$, then it is necessary to prescribe $D-r$ of the partial derivatives $\PD{\Theta}{\rchi}$ in order to be able to determine the remaining~$r$ derivatives from the PDE and the initial data. 

We close in remarking that this is a common feature in gauge theories. Also, further examples of how to apply the tools introduced here can be found in the Appendix A.2 and A.3 of~\cite{DAmbrosio:2022}.

\subsection{Second Order PDEs and the Kinetic Matrix}
The results of the previous subsection can be readily generalized to higher order PDEs. In particular, the Cauchy or initial value problem for a \textit{quasi-linear second order PDE} can be written as
\begin{align}\label{eq:2ndOrderIVP}
	\begin{cases}
		\displaystyle\sum_{i=1, i\leq j}^{D}\ \sum_{j=1}^{D} M^{(ij)}(x)\,\partial_i \partial_j \Psi + \sum_{i=1}^{d} N^{(i)}(x)\, \partial_i \Psi + L(x)\, \Psi + V(x) = 0\\
		\left.\Psi\right|_\S = F\\
		\left.n^\mu \partial_\mu\Psi\right|_\S = G
	\end{cases}\,.
\end{align}
Since now it is necessary to perform two integrations in order to determine $\Psi$, we have to specify two initial value functions on $\S$. As before, we have to specify $\Psi$ on $\S$, but we also have to say what the derivative of $\Psi$ in the direction orthogonal to $\S$ is. That is the meaning of $\left.n^\mu \partial_\mu \Psi\right|_\S$, where $n^\mu$ is the unit normal vector to $\S$.

By following the same ideas and steps as in the previous subsection, one can prove that the initial value problem~\eqref{eq:2ndOrderIVP} has a unique solution if and only if
\begin{align}\label{eq:2ndOrderSolvabilityCondition}
	\det\left(\sum_{i=1, i\leq j}^{D}\sum_{j=1}^{D} \tilde{M}^{(ij)}(x)\, n_{i}(x) \,n_{j}(x)\right) &\neq 0  &\text{for all } x\in\S\,.
\end{align}
This is the obvious analogue of equation~\eqref{eq:1stOrderSolvabilityCriterion} which we found for first order PDEs. In practice, it is often more convenient to write the initial value problem~\eqref{eq:2ndOrderIVP} in a slightly different and more intuitive form. First of all, we assume that we can pick out a time coordinate from $x^\mu = (x^{1}, \dots, x^{D})$. Our convention is to set $x^{D} = t$ and call it \textit{the} time coordinate. The coordinates $(x^{1}, \dots, x^{D-1})$ are consequently referred to as spatial coordinates. Furthermore, we define $\dot{\Psi} \ce \partial_t \Psi$ and we take the Cauchy surface $\S$ to be a~$t=const.$ surface. This latter choice in particular implies that $\left.n^\mu \partial_\mu \Psi\right|_\S = \left.\dot{\Psi}\right|_\S = G$. This is essentially the physicist's version of the mathematically more general formulation presented in the previous subsection. 

Given this split of spacetime into space and time, we can also divide the matrices $M^{(ij)}$ into matrices with time-time indices, space-time, and space-space indices. Concretely, the initial value problem can now be recast into the form
\begin{align}\label{eq:2ndOrderPDE}
	\begin{cases}
		\displaystyle\mathcal{K}(x)\, \ddot{\Psi} + \sum_{i=1}^{D-1}\mathcal{M}^{(i)}(x)\, \partial_i \dot{\Psi} + \sum_{i=1, i\leq j}^{D-1}\sum_{j=1}^{D-1} \mathcal{P}^{(ij)}(x)\, \partial_i \partial_j \Psi + \text{lower order terms } = 0\\
		\left.\Psi\right|_\S = F\\
		\left.\dot{\Psi}\right|_\S = G
	\end{cases}\,,
\end{align}
where $\mathcal{K}$ is the $N\times N$ \textit{kinetic matrix}, $\mathcal{M}^{(i)}$ are $D-1$ matrices of dimension $N\times N$ which we refer to as \textit{mixing matrices} (since they mix spatial and temporal derivatives of~$\Psi$, i.e., derivatives of the form $\partial_i \dot{\Psi}$), and finally we have the $\frac{D(D-1)}{2}$ matrices $\mathcal{P}^{(ij)}$ of dimension~$N\times N$, which we call \textit{potential matrices}. 

It is easy to see that in the coordinates we have chosen, the solvability condition~\eqref{eq:2ndOrderSolvabilityCondition} simply reads
\begin{align}
	\det\left(\mathcal{K}(x)\right) &\neq 0  &\text{for all } x\in\S\,.
\end{align}
Thus, we immediately see why the kinetic matrix is so important in determining the number of degrees of freedom propagated by second order PDEs: If $\mathcal{K}$ has full rank, it also has a non-vanishing determinant and we can thus solve for all $\Psi$'s, i.e., we can integrate the second order PDEs. If, however, $\mathcal{K}$ does \textit{not} have maximal rank, its determinant is also zero and the PDEs cannot be integrated. This means we are not able to determine all the fields $\Psi$ just from the initial data and the PDEs. More input is needed.

To see this, we simply notice that we can always perform linear operations (addition/subtraction of equations, multiplication by real numbers/functions) on our system of second order PDEs~\eqref{eq:2ndOrderPDE}. Using such operations, we can always bring the system into a form in which the kinetic matrix has a row-echelon form. We denote the row-echelon form of $\mathcal{K}$ by $\tilde{\mathcal{K}}$. If the rank of $\mathcal{K}$ is $r\ce \text{rank}(\mathcal{K})<N$, its row-echelon form will consist of $r$ non-trivial and linearly independent rows, while the last~$N-r$ rows are filled with zeros:
\begin{align}
	\tilde{\mathcal{K}} = \left(\begin{array}{cccccc}
	\tilde{\mathcal{K}}_{11} & \tilde{\mathcal{K}}_{12} & \tilde{\mathcal{K}}_{13} & \cdots & \tilde{\mathcal{K}}_{1(N-1)} & \tilde{\mathcal{K}}_{1N}	\\
	0 & \tilde{\mathcal{K}}_{22} & \tilde{\mathcal{K}}_{23} & \cdots & \tilde{\mathcal{K}}_{2(N-1)} & \tilde{\mathcal{K}}_{2N} \\
	\vdots & & \ddots & & \ddots & \vdots\\
	0 & 0 & 0 & \cdots & 0 & \tilde{\mathcal{K}}_{rN}\\ \hline
	0 & 0 & 0 & \cdots & 0 & 0\\
	\vdots & \vdots & \vdots & \cdots & \vdots & \vdots\\
	0 & 0 & 0 & \cdots & 0 & 0	
	\end{array}\right)\,.
\end{align}
 This means we can solve the PDEs for the first $r$ second order time derivatives of $\Psi$, namely for $\ddot{\Psi}_1, \ddot{\Psi}_2, \dots, \ddot{\Psi}_r$. However, there are no equations which determine $\ddot{\Psi}_{r+1}, \ddot{\Psi}_{r+2},\dots, \ddot{\Psi}_{N}$. Rather, it is possible that these second order time derivatives appear on the right hand side of the expressions one obtaines for $\ddot{\Psi}_1, \ddot{\Psi}_2, \dots, \ddot{\Psi}_r$. Thus, the initial data and the PDEs are not sufficient to determine $\ddot{\Psi}_1, \ddot{\Psi}_2, \dots, \ddot{\Psi}_r$, we also have to specify the fields $\Psi_{r+1}, \Psi_{r+2},\dots, \Psi_{N}$ everywhere by hand. 

The row-echelon form of $\mathcal{K}$ also tells us that we are not completely free in specifying the initial data. In fact, since the last $N-r$ rows of $\tilde{\mathcal{K}}$ are zero, this implies that the last $N-r$ equations of our system of PDEs contain at most first order time derivatives or no time derivatives at all. This follows from the fact that after performing the linear operations necessary to bring $\mathcal{K}$ into its row-echelon form, the matrices $\mathcal{M}^{(i)}$ will in general \textit{not} have their last $N-r$ rows filled with zeros. Thus, these equations represent constraint equations. The initial data have to respect these constraints.

Moreover, we have to study how these constraints behave under time evolution. Concretely, we are allowed to take time derivatives of our PDEs and this can potentially generate new constraint equations. The mechanism can be sketched as follows: 

We assume that the system~\eqref{eq:2ndOrderPDE} has been manipulated such that $\mathcal{K}$ is in row-echelon form. Its last $N-r$ equations are thus constraint equations. By taking time derivatives of these constraints, we obtain new equations of the form
\begin{align}\label{eq:TimeDerivativeConstraints}
	\sum_{i=1}^{D-1}\mathcal{M}^{(i)}\,\partial_i \ddot{\Psi} + \text{terms linear in $\partial_t \Psi$} = 0\,.
\end{align}
That is, we obtain second order time derivatives of $\Psi$. By taking spatial derivatives of the first $r$ equations of~\eqref{eq:2ndOrderPDE} we obtain schematically
\begin{align}\label{eq:SpatialDerivativesOfSystem}
	\mathcal{K}\, \partial_i\ddot{\Psi} + \text{terms linear in $\partial_t \Psi$} = 0\,.
\end{align}
Under the right circumstances, it can happen that equation~\eqref{eq:SpatialDerivativesOfSystem} can be used to eliminate~$\partial_i \ddot{\Psi}$ terms in equation~\eqref{eq:TimeDerivativeConstraints} and that we are left with new equations which are first order in time derivatives. Hence, it can happen that we encounter new constraint equations.

Even tough we have not worked out the precise conditions under which new constraints can appear, the argument we sketched above hinges on the presence of $\partial_i \ddot{\Psi}$ terms in the time derivative of the constraint equations. These terms can only appear if $\mathcal{M}^{(i)}$ are non-trivial matrices. If, however, the matrices $\mathcal{M}^{(i)}$ are absent, it is not possible to generate new constraints. This suggests the following approach: Find a field-redefinition such that the PDEs for the new fields contain no $\mathcal{M}^{(i)}$ matrices. Then, the number of dynamical degrees of freedom can be directly determined from the rank of $\mathcal{K}$. 

Notice that these field redefinitions can \textit{not} be linear transformations, since no linear transformation can ever eliminate a matrix. These redefinitions are highly non-trivial, but they can be found as follows:
Assuming that the PDEs~\eqref{eq:2ndOrderPDE} stem from an action principle, write the corresponding Lagrangian in the form
\begin{align}
	L = \dot{\Psi}^\transpose\,\mathcal{K}\,\dot{\Psi} + \sum_{i=1}^{D-1}\dot{\Psi}^\transpose\,\mathcal{M}^{(i)}\,\left(\partial_i \Psi\right) + \sum_{i=1, i\leq j}^{D-1}\sum_{j=1}^{D-1} \left(\partial_i\Psi\right)^\transpose\, \mathcal{P}^{(ij)} \left(\partial_j\Psi\right) + \dots
\end{align}
This can be written even more compactly using a single ``master matrix'', namely

\begin{align}
	L = \frac12\begin{pmatrix}
		\dot{\Psi} \\
		\partial_1 \Psi \\
		\partial_2 \Psi \\
		\vdots \\
		\partial_{d}\Psi
	\end{pmatrix}^\transpose
	\begin{pmatrix}
		2\mathcal{K} & \mathcal{M}^{(1)} & \mathcal{M}^{(2)} & \dots & \mathcal{M}^{(d)} \\
		\mathcal{M}^{(1)} & 2\mathcal{P}^{(11)} & \mathcal{P}^{(12)} & \cdots & \mathcal{P}^{(1d)} \\
		\mathcal{M}^{(2)} & \mathcal{P}^{(12)} & 2\mathcal{P}^{(22)} & \cdots & \mathcal{P}^{(2d)} \\
		\vdots & & & \cdots &  \vdots \\
		\mathcal{M}^{(d)} &\mathcal{P}^{(1d)} & \mathcal{P}^{(2d)} & \cdots & 2\mathcal{P}^{(dd)}
	\end{pmatrix}
	\begin{pmatrix}
		\dot{\Psi} \\
		\partial_1 \Psi \\
		\partial_2 \Psi \\
		\vdots \\
		\partial_{d}\Psi
	\end{pmatrix}
	+ \dots\,,
\end{align}
where we have introduced $d\ce D-1$ for brevity. Diagonalizing this master matrix is tantamount to finding a field redefinition such that the mixing matrices $\mathcal{M}^{(i)}$ are all absent. Moreover, the kinetic matrix is then automatically diagonal and the number of propagating degrees of freedom can be read off by counting the non-zero diagonal elements.

\normalsize

\newpage
\bibliographystyle{JHEP}
\bibliography{Bibliography}

\providecommand{\href}[2]{#2}\begingroup\raggedright\begin{thebibliography}{10}

\bibitem{Olmo:2011uz}
G.~J. Olmo, \emph{{Palatini Approach to Modified Gravity: f(R) Theories and
  Beyond}}, \href{https://doi.org/10.1142/S0218271811018925}{\emph{Int. J. Mod.
  Phys. D} {\bfseries 20} (2011) 413}
  [\href{https://arxiv.org/abs/1101.3864}{{\ttfamily 1101.3864}}].

\bibitem{Heisenberg:2018}
L.~Heisenberg, \emph{{A systematic approach to generalisations of General
  Relativity and their cosmological implications}},
  \href{https://doi.org/10.1016/j.physrep.2018.11.006}{\emph{Phys. Rept.}
  {\bfseries 796} (2019) 1} [\href{https://arxiv.org/abs/1807.01725}{{\ttfamily
  1807.01725}}].

\bibitem{BeltranJimenez:2019}
J.~Beltr\'an~Jim\'enez, L.~Heisenberg and T.~S. Koivisto, \emph{{The
  Geometrical Trinity of Gravity}},
  \href{https://doi.org/10.3390/universe5070173}{\emph{Universe} {\bfseries 5}
  (2019) 173} [\href{https://arxiv.org/abs/1903.06830}{{\ttfamily
  1903.06830}}].

\bibitem{BeltranJimenez:2017b}
J.~Beltran~Jimenez, L.~Heisenberg and T.~Koivisto, \emph{{Coincident General
  Relativity}},  \href{https://arxiv.org/abs/1710.03116}{{\ttfamily
  1710.03116}}.

\bibitem{BeltranJimenez:2019tme}
J.~Beltr\'an~Jim\'enez, L.~Heisenberg, T.~S. Koivisto and S.~Pekar,
  \emph{{Cosmology in $f(Q)$ geometry}},
  \href{https://doi.org/10.1103/PhysRevD.101.103507}{\emph{Phys. Rev. D}
  {\bfseries 101} (2020) 103507}
  [\href{https://arxiv.org/abs/1906.10027}{{\ttfamily 1906.10027}}].

\bibitem{DAmbrosio:2020c}
F.~D'Ambrosio, M.~Garg and L.~Heisenberg, \emph{{Non-linear extension of
  non-metricity scalar for MOND}},
  \href{https://doi.org/10.1016/j.physletb.2020.135970}{\emph{Phys. Lett. B}
  {\bfseries 811} (2020) 135970}
  [\href{https://arxiv.org/abs/2004.00888}{{\ttfamily 2004.00888}}].

\bibitem{Bajardi:2020}
F.~Bajardi, D.~Vernieri and S.~Capozziello, \emph{{Bouncing Cosmology in f(Q)
  Symmetric Teleparallel Gravity}},
  \href{https://doi.org/10.1140/epjp/s13360-020-00918-3}{\emph{Eur. Phys. J.
  Plus} {\bfseries 135} (2020) 912}
  [\href{https://arxiv.org/abs/2011.01248}{{\ttfamily 2011.01248}}].

\bibitem{Ayuso:2020}
I.~Ayuso, R.~Lazkoz and V.~Salzano, \emph{{Observational constraints on
  cosmological solutions of $f(Q)$ theories}},
  \href{https://doi.org/10.1103/PhysRevD.103.063505}{\emph{Phys. Rev. D}
  {\bfseries 103} (2021) 063505}
  [\href{https://arxiv.org/abs/2012.00046}{{\ttfamily 2012.00046}}].

\bibitem{Frusciante:2021}
N.~Frusciante, \emph{{Signatures of $f(Q)$-gravity in cosmology}},
  \href{https://doi.org/10.1103/PhysRevD.103.044021}{\emph{Phys. Rev. D}
  {\bfseries 103} (2021) 044021}
  [\href{https://arxiv.org/abs/2101.09242}{{\ttfamily 2101.09242}}].

\bibitem{Anagnostopoulos:2021}
F.~K. Anagnostopoulos, S.~Basilakos and E.~N. Saridakis, \emph{{First evidence
  that non-metricity f(Q) gravity could challenge \ensuremath{\Lambda}CDM}},
  \href{https://doi.org/10.1016/j.physletb.2021.136634}{\emph{Phys. Lett. B}
  {\bfseries 822} (2021) 136634}
  [\href{https://arxiv.org/abs/2104.15123}{{\ttfamily 2104.15123}}].

\bibitem{Atayde:2021}
L.~Atayde and N.~Frusciante, \emph{{Can $f(Q)$ gravity challenge
  $\Lambda$CDM?}},
  \href{https://doi.org/10.1103/PhysRevD.104.064052}{\emph{Phys. Rev. D}
  {\bfseries 104} (2021) 064052}
  [\href{https://arxiv.org/abs/2108.10832}{{\ttfamily 2108.10832}}].

\bibitem{DAmbrosio:2021b}
F.~D'Ambrosio, L.~Heisenberg and S.~Kuhn, \emph{{Revisiting cosmologies in
  teleparallelism}},
  \href{https://doi.org/10.1088/1361-6382/ac3f99}{\emph{Class. Quant. Grav.}
  {\bfseries 39} (2022) 025013}
  [\href{https://arxiv.org/abs/2109.04209}{{\ttfamily 2109.04209}}].

\bibitem{Capozziello:2022}
S.~Capozziello and R.~D'Agostino, \emph{{Model-independent reconstruction of
  f(Q) non-metric gravity}},
  \href{https://doi.org/10.1016/j.physletb.2022.137229}{\emph{Phys. Lett. B}
  {\bfseries 832} (2022) 137229}
  [\href{https://arxiv.org/abs/2204.01015}{{\ttfamily 2204.01015}}].

\bibitem{Dimakis:2022}
N.~Dimakis, A.~Paliathanasis, M.~Roumeliotis and T.~Christodoulakis,
  \emph{{FLRW solutions in f(Q) theory: The effect of using different
  connections}}, \href{https://doi.org/10.1103/PhysRevD.106.043509}{\emph{Phys.
  Rev. D} {\bfseries 106} (2022) 043509}
  [\href{https://arxiv.org/abs/2205.04680}{{\ttfamily 2205.04680}}].

\bibitem{Esposito:2022}
F.~Esposito, S.~Carloni and S.~Vignolo, \emph{{Bianchi type-I cosmological
  dynamics in $f(\mathcal{Q})$ gravity: a covariant approach}},
  \href{https://arxiv.org/abs/2207.14576}{{\ttfamily 2207.14576}}.

\bibitem{Zhao:2021}
D.~Zhao, \emph{{Covariant formulation of f(Q) theory}},
  \href{https://doi.org/10.1140/epjc/s10052-022-10266-4}{\emph{Eur. Phys. J. C}
  {\bfseries 82} (2022) 303}
  [\href{https://arxiv.org/abs/2104.02483}{{\ttfamily 2104.02483}}].

\bibitem{Lin:2021}
R.-H. Lin and X.-H. Zhai, \emph{{Spherically symmetric configuration in $f(Q)$
  gravity}}, \href{https://doi.org/10.1103/PhysRevD.103.124001}{\emph{Phys.
  Rev. D} {\bfseries 103} (2021) 124001}
  [\href{https://arxiv.org/abs/2105.01484}{{\ttfamily 2105.01484}}].

\bibitem{DAmbrosio:2021}
F.~D'Ambrosio, S.~D.~B. Fell, L.~Heisenberg and S.~Kuhn, \emph{{Black holes in
  f(Q) gravity}},
  \href{https://doi.org/10.1103/PhysRevD.105.024042}{\emph{Phys. Rev. D}
  {\bfseries 105} (2022) 024042}
  [\href{https://arxiv.org/abs/2109.03174}{{\ttfamily 2109.03174}}].

\bibitem{Banerjee:2021}
A.~Banerjee, A.~Pradhan, T.~Tangphati and F.~Rahaman, \emph{{Wormhole
  geometries in $f(Q)$ gravity and the energy conditions}},
  \href{https://doi.org/10.1140/epjc/s10052-021-09854-7}{\emph{Eur. Phys. J. C}
  {\bfseries 81} (2021) 1031}
  [\href{https://arxiv.org/abs/2109.15105}{{\ttfamily 2109.15105}}].

\bibitem{Wang:2021}
W.~Wang, H.~Chen and T.~Katsuragawa, \emph{{Static and spherically symmetric
  solutions in f(Q) gravity}},
  \href{https://doi.org/10.1103/PhysRevD.105.024060}{\emph{Phys. Rev. D}
  {\bfseries 105} (2022) 024060}
  [\href{https://arxiv.org/abs/2110.13565}{{\ttfamily 2110.13565}}].

\bibitem{Parsaei:2022}
F.~Parsaei, S.~Rastgoo and P.~K. Sahoo, \emph{{Wormhole in $f(Q)$ gravity}},
  \href{https://arxiv.org/abs/2203.06374}{{\ttfamily 2203.06374}}.

\bibitem{Maurya:2022}
S.~K. Maurya, K.~Newton~Singh, S.~V. Lohakare and B.~Mishra, \emph{{Anisotropic
  Strange Star Model Beyond Standard Maximum Mass Limit by Gravitational
  Decoupling in $f(Q)$ Gravity}},
  \href{https://arxiv.org/abs/2208.04735}{{\ttfamily 2208.04735}}.

\bibitem{Hu:2022}
K.~Hu, T.~Katsuragawa and T.~Qiu, \emph{{ADM formulation and Hamiltonian
  analysis of f(Q) gravity}},
  \href{https://doi.org/10.1103/PhysRevD.106.044025}{\emph{Phys. Rev. D}
  {\bfseries 106} (2022) 044025}
  [\href{https://arxiv.org/abs/2204.12826}{{\ttfamily 2204.12826}}].

\bibitem{Nester:1998}
J.~M. Nester and H.-J. Yo, \emph{{Symmetric teleparallel general relativity}},
  {\emph{Chin. J. Phys.} {\bfseries 37} (1999) 113}
  [\href{https://arxiv.org/abs/gr-qc/9809049}{{\ttfamily gr-qc/9809049}}].

\bibitem{BeltranJimenez:2018}
J.~Beltr{\'a}n~Jim{\'e}nez, L.~Heisenberg and T.~S. Koivisto,
  \emph{{Teleparallel Palatini theories}},
  \href{https://arxiv.org/abs/1803.10185}{{\ttfamily 1803.10185}}.

\bibitem{DAmbrosio:2020}
F.~D'Ambrosio, M.~Garg, L.~Heisenberg and S.~Zentarra, \emph{{ADM formulation
  and Hamiltonian analysis of Coincident General Relativity}},
  \href{https://arxiv.org/abs/2007.03261}{{\ttfamily 2007.03261}}.

\bibitem{DAmbrosio:2020b}
F.~D'Ambrosio and L.~Heisenberg, \emph{{Classification of primary constraints
  of quadratic non-metricity theories of gravity}},
  \href{https://doi.org/10.1007/JHEP02(2021)170}{\emph{JHEP} {\bfseries 02}
  (2021) 170} [\href{https://arxiv.org/abs/2007.05064}{{\ttfamily
  2007.05064}}].

\bibitem{Dirac:1950}
P.~A.~M. Dirac, \emph{Generalized hamiltonian dynamics},
  \href{https://doi.org/10.4153/CJM-1950-012-1}{\emph{Canadian Journal of
  Mathematics} {\bfseries 2} (1950) 129}.

\bibitem{Bergmann:1951}
J.~L. Anderson and P.~G. Bergmann, \emph{Constraints in covariant field
  theories}, \href{https://doi.org/10.1103/PhysRev.83.1018}{\emph{Phys. Rev.}
  {\bfseries 83} (1951) 1018}.

\bibitem{DiracBook}
P.~A.~M. Dirac, \emph{{Lectures on Quantum Mechanics}}. Dover Publications,
  1964.

\bibitem{Ferraro:2016}
R.~Ferraro and M.~J. Guzm\'an, \emph{{Hamiltonian formulation of teleparallel
  gravity}}, \href{https://doi.org/10.1103/PhysRevD.94.104045}{\emph{Phys. Rev.
  D} {\bfseries 94} (2016) 104045}
  [\href{https://arxiv.org/abs/1609.06766}{{\ttfamily 1609.06766}}].

\bibitem{Blixt:2018}
D.~Blixt, M.~Hohmann and C.~Pfeifer, \emph{{Hamiltonian and primary constraints
  of new general relativity}},
  \href{https://doi.org/10.1103/PhysRevD.99.084025}{\emph{Phys. Rev. D}
  {\bfseries 99} (2019) 084025}
  [\href{https://arxiv.org/abs/1811.11137}{{\ttfamily 1811.11137}}].

\bibitem{Blixt:2019}
D.~Blixt, M.~Hohmann, M.~Kr\v{s}\v{s}\'ak and C.~Pfeifer, \emph{{Hamiltonian
  Analysis In New General Relativity}},  in \emph{{15th Marcel Grossmann
  Meeting on Recent Developments in Theoretical and Experimental General
  Relativity, Astrophysics, and Relativistic Field Theories}}, 5, 2019,
  \href{https://arxiv.org/abs/1905.11919}{{\ttfamily 1905.11919}},
  \href{https://doi.org/10.1142/9789811258251_0038}{DOI}.

\bibitem{Blixt:2020}
D.~Blixt, M.-J. Guzm\'an, M.~Hohmann and C.~Pfeifer, \emph{{Review of the
  Hamiltonian analysis in teleparallel gravity}},
  \href{https://doi.org/10.1142/S0219887821300051}{\emph{Int. J. Geom. Meth.
  Mod. Phys.} {\bfseries 18} (2021) 2130005}
  [\href{https://arxiv.org/abs/2012.09180}{{\ttfamily 2012.09180}}].

\bibitem{Li:2011}
M.~Li, R.-X. Miao and Y.-G. Miao, \emph{{Degrees of freedom of $f(T)$
  gravity}}, \href{https://doi.org/10.1007/JHEP07(2011)108}{\emph{JHEP}
  {\bfseries 07} (2011) 108} [\href{https://arxiv.org/abs/1105.5934}{{\ttfamily
  1105.5934}}].

\bibitem{Blagojevic:2020}
M.~Blagojevi\'c and J.~M. Nester, \emph{{Local symmetries and physical degrees
  of freedom in $f(T)$ gravity: a Dirac Hamiltonian constraint analysis}},
  \href{https://doi.org/10.1103/PhysRevD.102.064025}{\emph{Phys. Rev. D}
  {\bfseries 102} (2020) 064025}
  [\href{https://arxiv.org/abs/2006.15303}{{\ttfamily 2006.15303}}].

\bibitem{Ferraro:2018}
R.~Ferraro and M.~J. Guzm\'an, \emph{{Hamiltonian formalism for f(T) gravity}},
  \href{https://doi.org/10.1103/PhysRevD.97.104028}{\emph{Phys. Rev. D}
  {\bfseries 97} (2018) 104028}
  [\href{https://arxiv.org/abs/1802.02130}{{\ttfamily 1802.02130}}].

\bibitem{HenneauxBook}
M.~Henneaux and C.~Teitelboim, \emph{{Quantization of gauge systems}}. 1992.

\bibitem{Wipf:1993}
A.~W. Wipf, \emph{{Hamilton's formalism for systems with constraints}},
  \href{https://doi.org/10.1007/3-540-58339-4\_14}{\emph{Lect. Notes Phys.}
  {\bfseries 434} (1994) 22}
  [\href{https://arxiv.org/abs/hep-th/9312078}{{\ttfamily hep-th/9312078}}].

\bibitem{SundermeyerBook}
K.~Sundermeyer, \emph{Constrained Dynamics}, Lecture Notes in Physics.
  Springer, 1982.

\bibitem{BeltranJimenez:2020}
J.~Beltr\'an~Jim\'enez, L.~Heisenberg and T.~Koivisto, \emph{{The coupling of
  matter and spacetime geometry}},
  \href{https://doi.org/10.1088/1361-6382/aba31b}{\emph{Class. Quant. Grav.}
  {\bfseries 37} (2020) 195013}
  [\href{https://arxiv.org/abs/2004.04606}{{\ttfamily 2004.04606}}].

\bibitem{MiersemannBook}
E.~Miersemann, \emph{{Partial Differential Equations}}. 2014.

\bibitem{DAmbrosio:2022}
F.~D'Ambrosio, S.~D.~B. Fell, L.~Heisenberg, D.~Maibach, S.~Zentarra and
  J.~Zosso, \emph{{Gravitational Waves in Full, Non-Linear General
  Relativity}},  \href{https://arxiv.org/abs/2201.11634}{{\ttfamily
  2201.11634}}.

\end{thebibliography}\endgroup
\end{document}